\documentclass[preprint]{aastex}
\usepackage{amssymb,amsmath,graphicx,natbib,upgreek,float}

\begin{document}

\title{The Shock Dynamics of Heterogeneous YSO Jets: 3-D Simulations Meet Multi-Epoch Observations}
\author{E. C. Hansen and A. Frank}
\affil{Department of Physics and Astronomy, University of Rochester, Rochester, NY 14627-0171, USA}
\author{P. Hartigan}
\affil{Department of Physics and Astronomy, Rice University, 6100 S. Main, Houston, TX 77521-1892, USA}
\author{S. V. Lebedev}
\affil{Blackett Laboratory, Imperial College London, Prince Consort Road, London SW7 2BW, UK}

\begin{abstract}
High resolution observations of Young Stellar Object (YSO) jets show them to be composed of many small-scale knots or clumps.
In this paper we report results of 3-D numerical simulations designed to study how such clumps interact and create morphologies and kinematic patterns seen in emission line observations.
Our simulations focus on clump scale dynamics by imposing velocity differences between spherical, over-dense regions which then lead to the formation of bow shocks as faster clumps overtake slower material.
We show that much of the spatial structure apparent in emission line images of jets arises from the dynamics and interactions of these bow shocks.
Our simulations show a variety of time-dependent features, including bright knots associated with Mach stems where the shocks intersect, a ``frothy'' emission structure that arises from the presence of the Non-linear Thin Shell Instability (NTSI) along the surfaces of the bow shocks, and the merging and fragmentation of clumps.
Our simulations use a new non-equilibrium cooling method to produce synthetic emission maps in H$\alpha$ and [S II].
These are directly compared with multi-epoch Hubble Space Telescope (HST) observations of Herbig-Haro (HH) jets.
We find excellent agreement between features seen in the simulations and the observations in terms of both proper motion and morphologies.
Thus we conclude that YSO jets may be dominated by heterogeneous structures and that interactions between these structures and the shocks they produce can account for many details of YSO jet evolution. 
\end{abstract}


\section{Introduction}
\label{sec:intro}
Herbig-Haro (HH) objects, associated with jets from young stellar objects (YSOs), are an integral part of the star formation process \citep[for an overview, see][]{Frank14}.
These jets carry away angular momentum which is necessary for the central object to accrete material, and they interact with the natal molecular cloud material as they travel outwards.
Since the outflows are so closely linked to the accretion process, it follows that time variability in the accreting disk produces variability in the outflowing jet \citep{Cabrit90,Hartigan95}.
During the jet launch process, which is currently understood to be magneto-centrifugally driven \citep{Blandford82,Uchida85}, MHD instabilities can disrupt the jet beam leading to considerable spatial variability.
These instabilities have been seen in both simulations \citep{Huarte-Espinosa12} and laboratory plasma experiments \citep{Ciardi09,Ciardi07,Lebedev05}.

Both variability at the source, or instabilities active during the launching, will lead to ``clumpiness'' within the jet beam.
Such clumpy jets have been studied by a number of authors looking to understand the details of heterogeneous flow interactions \citep[e.g.][]{Raga02ApJ,Hartigan05,Beck07}.
In particular the work of \citet{Yirak12} explored the dynamics of clumps moving with a range of velocities relative to a background ``jet beam''.
These studies showed how clumps interact with the jet beam forming secondary bow shocks which may then interact with each other.
These secondary bow shocks can have different velocities due to the pulsation and precession of the outflowing jet.
There is evidence of such interactions in the high resolution multi-epoch observations as they show that different components of the gas within HH objects can have different velocities \citep{Hartigan11}.
In this paper, we focus in detail on the interactions between such clumps, the bow shocks they drive, and the emission patterns that emerge from the dynamics.

We note that exploring clump dynamics means exploring clump and bow shock instabilities.
Some of the observed emission features in clumpy jets are caused by hydrodynamic instabilities in both the clumps and their bow shocks.
The Rayleigh-Taylor (RT), Kelvin-Helmholtz (KH), and Non-linear Thin Shell Instability (NTSI) all have a role to play in the dynamics of these outflows and thus the patterns found in HH jet emission maps.

The relevant observations which motivate this study come from high resolution multi-epoch observations of a number of HH jets (\citet{Hartigan11}, and references therein).
These observations show time-dependent structures which imply heterogeneous density distributions within jet beams ($R_{clump} < R_{jet}$).
For example, bright knots of H$\alpha$ emission are seen both within the jets and at the jet bow shocks (or internal working surfaces).
Some of these H$\alpha$ knots are located at the intersection points between what appear to be separate bow shocks.
These features raise the possibility that the bright emission is due to the formation of a Mach stem at the intersection of bow shocks from individual clumps.
When bow shocks intersect at an angle at, or above, a certain minimum critical value, a third shock (the Mach stem) can form.
Since Mach stems are planar shocks, gas that crosses it will be heated to higher temperatures compared to the oblique bow shock.
As a result, enhanced emission will be produced relative to that from a single bow shock \citep{Hansen15b}.
Such intersections can also exhibit motion that is lateral to the overall flow which appears to be seen in the observations of HH objects such as HH 34. 

In this paper we attempt to recover behavior seen in HH jets such as that described above, through the use of high resolution 3-D radiative hydrodynamic numerical simulations.
We note that H$\alpha$ emission in outflows typically marks shock fronts while regions of [S II] emission are observed in cooling regions behind the shocks \citep{Heathcote96}.
Several groups have studied the emission produced by variable velocity jet models \citep[e.g.][]{Kajdic06,deColle06,Raga07}, and recent high resolution MHD simulations by \citet{Hansen15a} have explored how these variable velocity jet models can produce internal structure and their resulting H$\alpha$ and [S II] emission patterns.
As in \citet{Hansen15a}, the numerical simulations in this work contain accurate non-equilibrium cooling which enables us to make synthetic H$\alpha$ and [S II] emission maps.
These allow direct comparisons with HST images of HH objects. 

The structure of the paper is as follows: in Section~\ref{sec:methods}, we present our numerical methods as well as our initial conditions for the various models.
In Section~\ref{sec:theory} we discuss the various instabilities and timescales relevant to our simulations.
We also give a brief overview of the theory of Mach stem formation and how bow shock intersections affect emission.
Section~\ref{sec:results} contains our simulation results and our interpretations.
Finally, we present our conclusions in Section~\ref{sec:conc}.

\section{Methods}
\label{sec:methods}
The simulations were carried out using AstroBEAR, a highly parallelized adaptive mesh refinement (AMR) multi-physics code.
See \cite{Cunningham09,Carroll12} for a detailed explanation of how AMR is implemented.
More details of the code can also be found at https://astrobear.pas.rochester.edu/trac/.
Here we provide a brief overview of the physics implemented for this study.
The code solves the 3-D Euler equations of fluid dynamics with non-equilibrium cooling:

\begin{subequations}\label{group1}
\begin{gather}
    \frac{\partial \rho}{\partial t} + \boldsymbol{\nabla} \cdot \rho \boldsymbol{v} = 0 ,\ \label{1a}\\[\jot]
    \frac{\partial \rho \boldsymbol{v}}{\partial t} + \boldsymbol{\nabla} \cdot (\rho \boldsymbol{v} \boldsymbol{v} + P\boldsymbol{I}) = 0 ,\ \label{1b}\\[\jot]
    \frac{\partial E}{\partial t} + \boldsymbol{\nabla} \cdot ((E + P) \boldsymbol{v}) = -L ,\ \label{1c}\\[\jot]
    \frac{\partial n_i}{\partial t} + \boldsymbol{\nabla} \cdot n_i \boldsymbol{v} = C_i ,\ \label{1d}
\end{gather}
\end{subequations}
where $\rho$ is the mass density, $\boldsymbol{v}$ is the velocity, $P$ is the thermal pressure, $\boldsymbol{I}$ is the identity matrix, and $E$ is the total energy such that $E = \frac{1}{\gamma - 1} P + \frac{1}{2}\rho v^2$ (with $\gamma = \frac{5}{3}$ for an ideal gas).
$L$ is the cooling source term which will be a function of number density, temperature and ionization.
$n_i$ is the number density of species $i$, and $C_i$ is the sum of the ionization and recombination processes for species $i$.

The equations above represent the conservation of mass \eqref{1a}, momentum \eqref{1b}, and energy \eqref{1c}.
Equation \eqref{1d} represents the evolution of the number densities of the different atomic species tracked within the code.

The cooling source term on the right hand side of equation \eqref{1c} is implemented in AstroBEAR through calculations using the ionization rate equations.
The rates are then used to determine the ionization and recombination energy losses from both hydrogen and helium.
Cooling from metal excitation is calculated from one of two different tables depending on the temperature.
Below $10^4 K$, we use a new cooling table which uses strong charge exchange cross sections to lock the ratios of NII/NI and OII/OI to HII/HI, and it solves multilevel atom models to derive volume emission cooling terms.
Above $10^4 K$, we use a modified version of the Dalgarno \& McCray cooling curve (\citet{Dalgarno72}) where we subtract the contributions from H and He, leaving only the metal components.
A more detailed description of the cooling table and other cooling processes relevant to radiative shocks will be given in a future paper (\citet[in preparation]{Hansen16}).

Using equation \eqref{1d} allows us to keep track of the number densities of the neutral and ionized species.
There is a total of 8 species tracked in the code: neutral and ionized hydrogen, neutral, ionized, and doubly ionized helium, and SII, SIII, and SIV.
Tracking the hydrogen and helium species allows us to track their ionization fractions and thus the electron number density of the gas which is required for generating synthetic emission maps.
Below $10^4 K$, all S is assumed to be SII, and above $10^4 K$, the ionization and recombination rates are used to track the amount of SIII and SIV.
This was employed to more accurately track the ionization state of S and hence produce more accurate [S II] maps.

Calculating the H$\alpha$ emission accurately is crucial to comparing these simulations with observations.
In AstroBEAR, the H$\alpha$ routine is dependent on electron number density, temperature, ionization fraction, and hydrogen number density.
There are two main components to the emission: a recombination term and a collisional excitation term.

We include excitations to levels 3, 4, and 5 using effective collision stengths from \citet{Anderson00} and \citet{Anderson02}.
The recombination term is important when the gas is highly ionized at relatively low temperatures ($<$ 10,000 K).
When ionized hydrogen is recombining, there is a nonzero probability that it will go through the H$\alpha$ transition.
Recombination coefficients are taken from \citet{Verner96}, and they are valid above 3000 K.

In order to generate a synthetic emission map, the gas is assumed to be optically thin, and the emission is summed along a line of sight.
The emission maps in this paper show [S II] (red) as well as H$\alpha$ (green), and the ratio [S II]/H$\alpha$ determines the color.
We can rotate and/or incline the simulation data to create projections along different lines of sight (see Figures~\ref{fig:3-clumptop} and \ref{fig:3-clumpside}).

\subsection{Models}
\label{subsec:models}
We conduct three different classes of simulations in order to articulate basic modalities of behavior.
There are 2-clump runs at high resolution, 3-clump runs at moderate resolution, and a multi-clump run at moderate resolution.
All simulations consist of the clumps embedded in a moving ambient.
The ambient velocity is constant such that the simulation is always in the reference frame of a clump moving with a Mach number of 7.

The 3-clump and multi-clump runs use 3 levels of AMR while the 2-clump runs use 4 levels which leads to effective resolutions of 40 and 80 cells per clump radius respectively.
One clump radius is equal to 10 AU.
All simulations use open boundary conditions except for the top which maintains inflow conditions equal to the initialized ambient medium (i.e. the background flow).
The ambient density, temperature, and ionization fraction are set at 10\textsuperscript{3} cm\textsuperscript{-3}, 1250 K, and 0.01 respectively.
The clumps are initialized with a density of 5 x 10\textsuperscript{5} cm\textsuperscript{-3} and are set in pressure equilibrium with the background gas.
Since the clumps are dense and cold, they are given a more realistic initial ionization fraction of 10\textsuperscript{-5}.
All simulations ran for approximately 75 years.

\begin{deluxetable}{cccc}
\tablecolumns{4}
\tablewidth{0pt}
\tablecaption{Simulation parameters for 2-clump and 3-clump models \label{tab:sims}}
\tablehead{\colhead{Model} & \colhead{Separation ($r_{clump}$)} & \colhead{Mach numbers of clump} & \colhead{Angle $a$ (deg)}}
\startdata
A & 2.5 & 10, 7 & --- \\
B & 4.5 & 10, 7 & --- \\
C & 6.5 & 10, 7 & --- \\
D & 2.5 & 15, 7 & --- \\
E & 4.5 & 15, 7 & --- \\
F & 6.5 & 15, 7 & --- \\
G & 2.5 & 15, 10, 7 & 30 \\
H & 2.5 & 15, 10, 7 & 60 \\
I & 2.5 & 15, 10, 7 & 90 \\
J & 4.5 & 15, 10, 7 & 30 \\
K & 4.5 & 15, 10, 7 & 60 \\
L & 4.5 & 15, 10, 7 & 90 \\
\enddata
\end{deluxetable}

\begin{figure}
\includegraphics[width=0.5\textwidth]{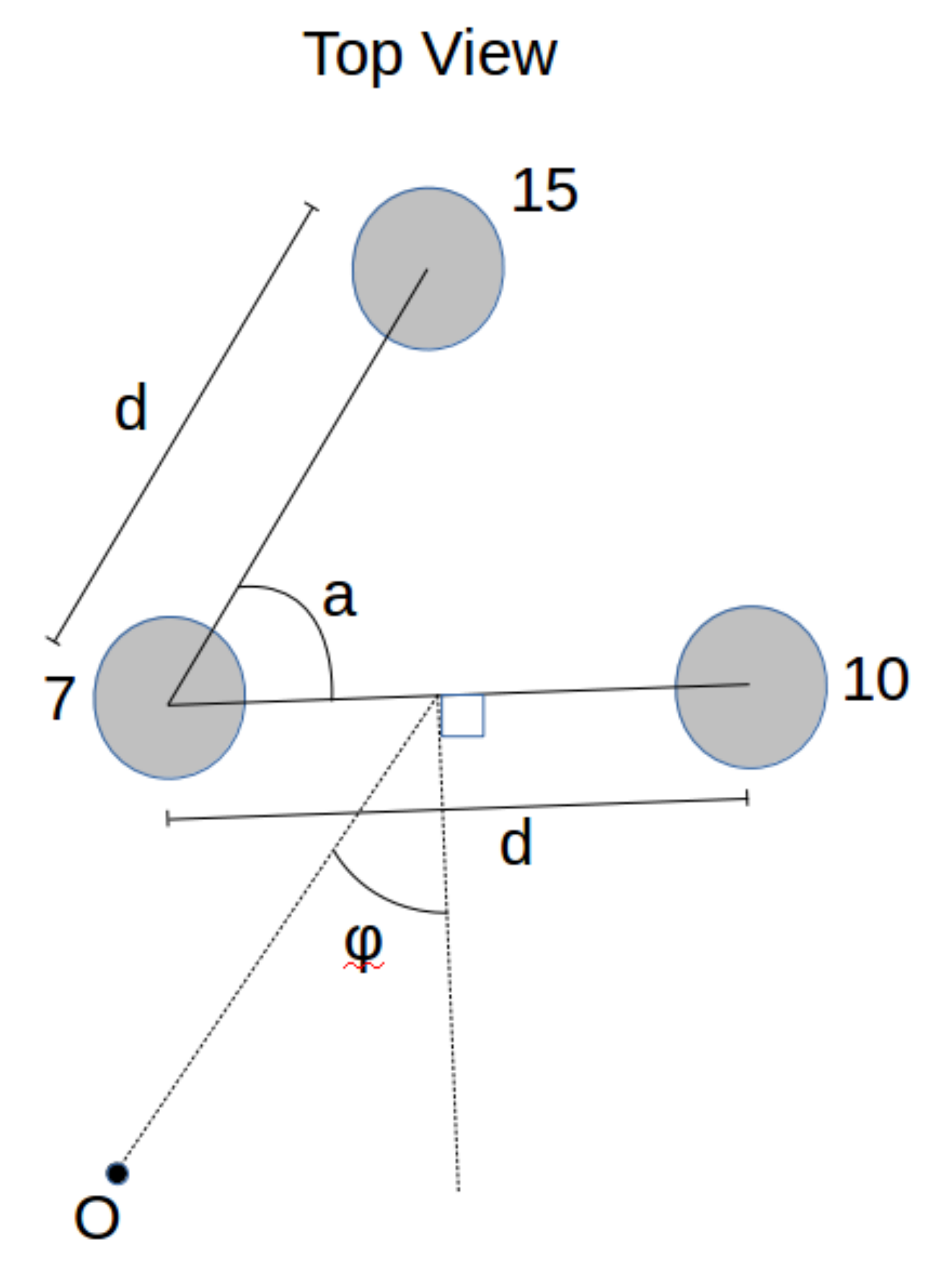}
\caption{Schematic of the 3-clump set up (top view).
Each clump is labeled by its Mach number.
The separation distance $d$ and angle $a$ are parameters that are varied from run to run.
The angle $\upvarphi$ is the rotation angle at which a line of sight is drawn from the observer at the point labled $O$.}
\label{fig:3-clumptop}
\end{figure}

For the 2-clump models, we ran 6 simulations varying the clump separation distance $d$ and the clump velocities.
Table~\ref{tab:sims} lists the varying simulation parameters for the 2-clump models (A - F).
The clump separation is in units of clump radii (10 AU), and the clump velocities are given as Mach numbers with respect to the ambient sound speed.

\begin{figure}
\includegraphics[width=0.5\textwidth]{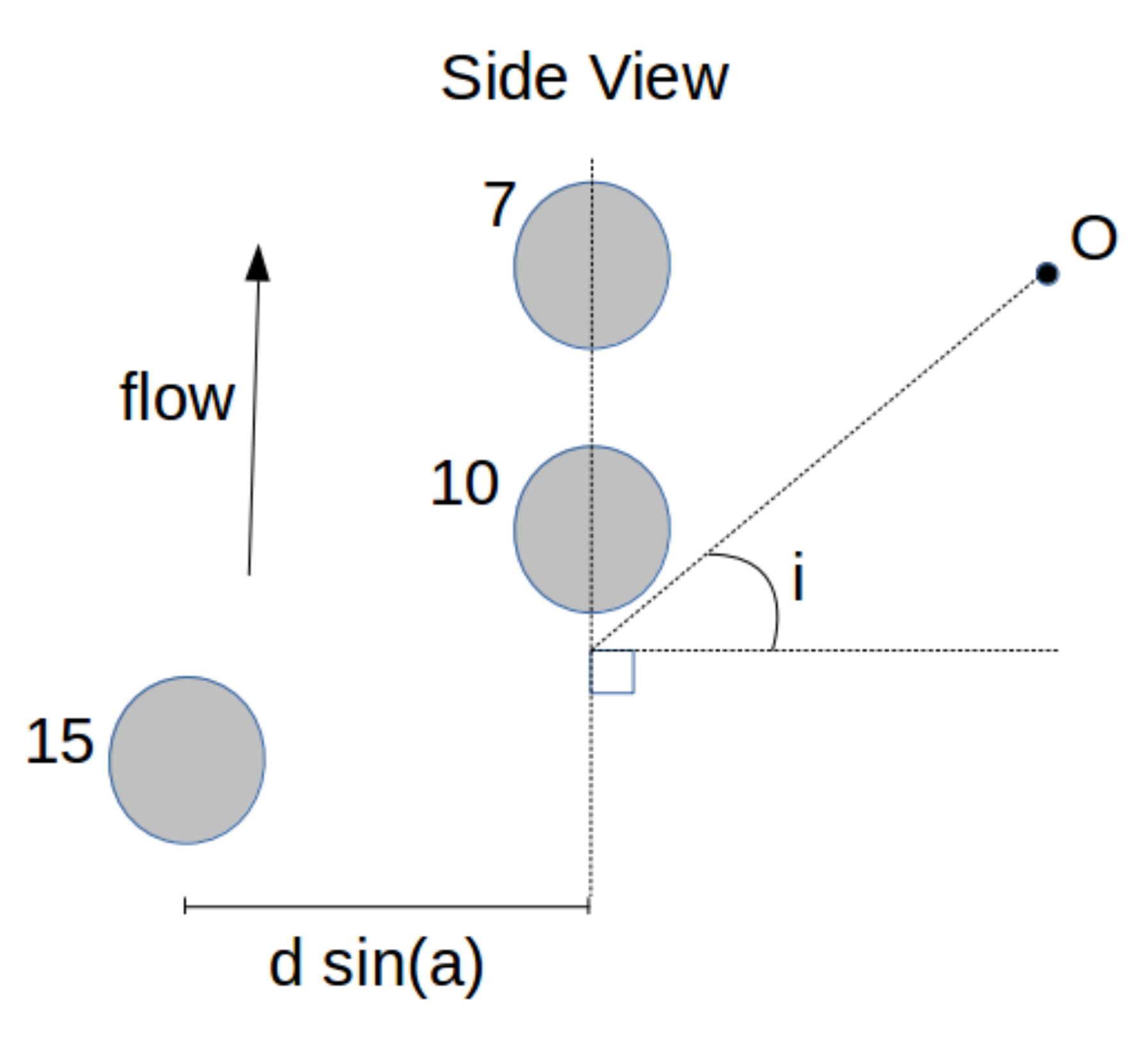}
\caption{Schematic of the 3-clump set up (side view).
All labels are the same as in Figure~\ref{fig:3-clumptop}, and angle $i$ refers to the inclination angle to the observer.}
\label{fig:3-clumpside}
\end{figure}

For the 3-clump models, each clump is moving at a different velocity: M = 15, 10, and 7.
We vary the separation distance between the clumps, and the angular position of the M = 15 clump.
Figures~\ref{fig:3-clumptop} and \ref{fig:3-clumpside} show simple schematics of this set-up, and Table~\ref{tab:sims} also lists the defining parameters for these models (G - L).. 
The figures also show the angles at which a line of sight is drawn from an observer, and the optically thin emission is summed along this line of sight to produce projected emission maps at different viewing angles.

The multi-clump run consists of 10 clumps positioned randomly on a lattice within the computational domain and with random velocities between M = 7 and M = 15.
Their x and z positions are fixed such that the clumps form a 3-4-3 lattice structure when viewed from above, and their y positions depend on their velocities in such a way that each clump will remain on the grid during the simulation.
The purpose of this run was to explore the formation of global bow shock patterns (``curtains'' or ``sheets'') as seen in objects such as HH 2 \citep{Hartigan11}.

All of the aforementioned densities, temperatures, velocities, and length scale are within the commonly accepted ranges for HH jets (see \cite{Frank14}, and refs. therein).

\section{Theory}
\label{sec:theory}
The complex morphologies observed in HH objects occur due to a variety of fluid dynamical processes such as Mach stem formation at bow intersection points and instabilities within the clumps and bow shocks.
There are observational and numerical challenges to studying these processes via simulations which seek to focus on offset clumps moving past each other in 3-D.
Thus, we begin by considering features of the simpler problem of two interacting stationary bows.
This problem was studied via 2-D numerical simulations by \citet{Hansen15b} whose goal was to determine the minimum critical angle for Mach stem formation.
We review this problem in the first subsection below, and in the second subsection, we describe the relevant hydrodynamic instabilities and how they should affect the observed flow and emission.

\subsection{Mach Stem Formation}
\label{subsec:machstem}
\begin{figure}
\includegraphics[width=0.5\textwidth]{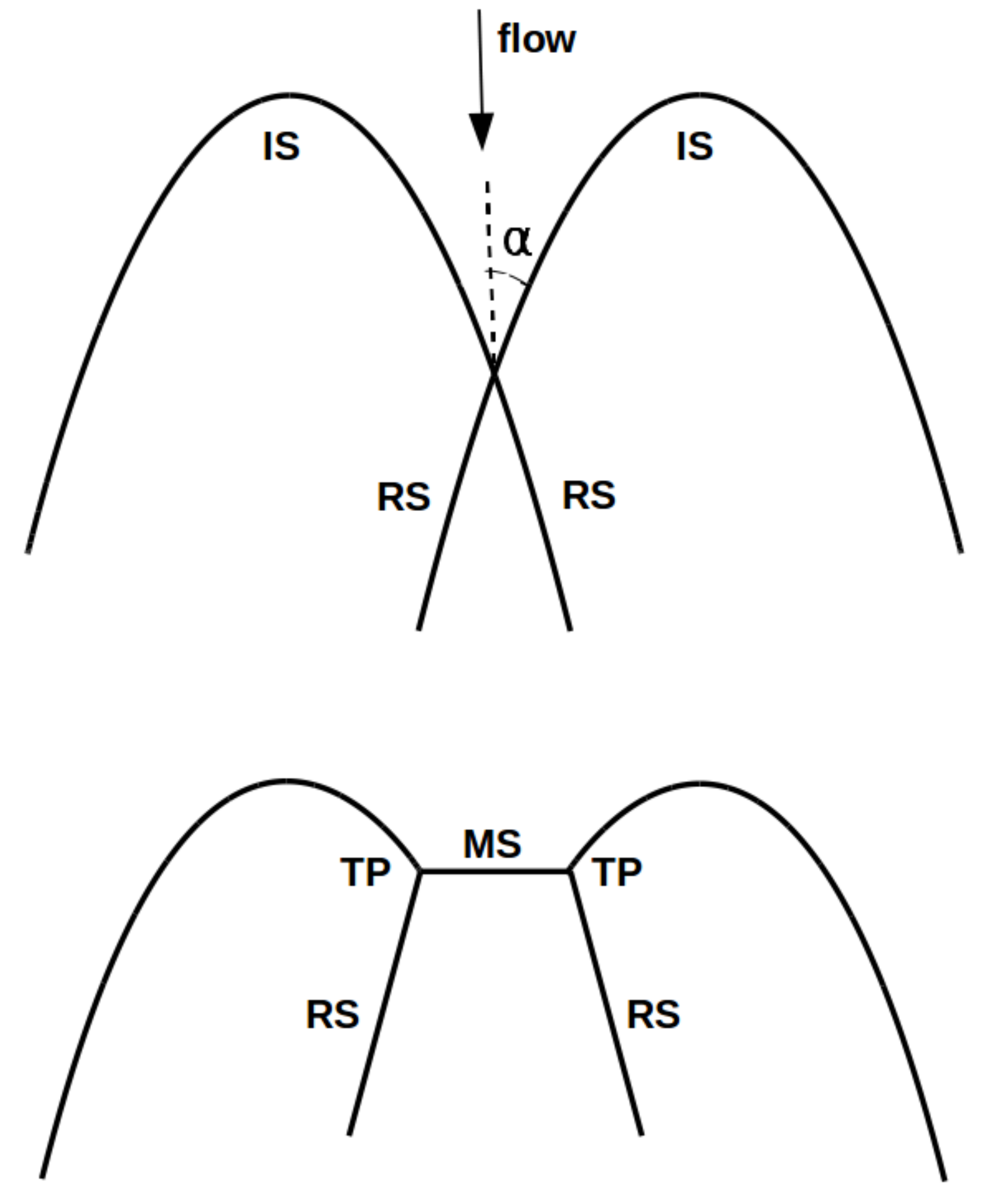}
\caption{Diagram of intersecting bow shocks with (\emph{bottom}) and without (\emph{top}) a Mach stem.
The acronyms are as follows: ``IS'' = incident shock (bow shock), ``RS'' = reflected shock, ``MS'' = Mach stem, ``TP'' = triple point.
Also shown is the direction of the flow and the included angle $\alpha$.}
\label{fig:reflect}
\end{figure}

When two bow shocks intersect, they reflect off of each other forming symmetric reflected oblique shocks; this is known as regular reflection (A description of Mach stem formation, also known as Mach reflection, can be found in several texts \citep[e.g.][]{BenDor07,Courant99,Landau87}).
Without a Mach stem, a gas parcel passes through the incident bow shock and then through the reflected shock.
The gas is deflected by the shocks in such a way that it recovers its initial propagation direction.
However, once the intersection of the bow shocks reaches a critical angle, (determined by the jump conditions for the colliding shocks), a gas parcel can no longer pass through both shocks and maintain its original direction of flow.
Therefore, a planar shock known as a Mach stem is formed so that the gas can continue to flow downstream in the same direction as the initial pre-shock flow.
The Mach stem extends from each bow shock at positions known as triple points representing the new positions where the bow shocks reflect (see Figure~\ref{fig:reflect}).

In both Mach reflection and regular reflection, the intersection point of two bow shocks will exhibit enhanced emission.
With Mach reflection, the Mach stem is stronger than the neighboring bow shock because it is a planar shock.
Hence, the Mach stem will heat the gas to a higher temperature and result in stronger H$\alpha$ emission.
In regular reflection, the gas near the intersection point will travel through two shocks: the incident shock and the reflected shock.
Therefore, the gas is shock-heated twice and will again result in brighter H$\alpha$ emission.
In our simulations, this region of bright emission behind the intersection point appears as a ``wedge''.

For the simulations that follow it is useful to consider the conditions under which a Mach stem should form.
For strong shocks, the critical angle $\alpha_c$ (which is a minimum for Mach stem formation) is only dependent on the ratio of specific heats $\gamma$.
A simple derivation of $\alpha_c$ leads to the following approximate formula \citep{deRosa92,Courant99}:
\begin{equation}
    \alpha_c = \arcsin(\frac{1}{\gamma}) .\ \label{2}
\end{equation}
In order to approximate strong cooling, $\gamma$ approaches 1, and under these conditions, the critical angle $\alpha_c$ reaches its limiting value of 90$^{\circ}$.
Since the flows in simulations will be strongly radiative, we can see that the angle between the intersecting bow shocks will need to be large in order to allow for the formation of a Mach stem.

It is difficult to initialize the intersection angles in our simulations, so we vary the intersection angles by varying the clump separation distances $d$.
Smaller clump separations imply larger bow shock intersection angles (as $d$ approaches 0, $\alpha$ approaches 90$^{\circ}$).
Thus the critical separation distance for Mach stem formation should be quite small (compared to the clump radius) when strong cooling is present.
Note, however, that these conditions which tell us when to expect a Mach stem also assume the shocks have smooth profiles which do not change in time.
As we will see, this is not often the case.

For values of $\gamma \simeq 1$, a bow shock shape will become highly time-dependent due to the NTSI which we will describe further in the next subsection \citep{Vishniac94}.
This bow shock instability can both hinder and support Mach stem formation at any separation distance.
In other words, when corrugated bow shocks intersect, their intersection angle is constantly changing, hence Mach stem formation will no longer depend on their separation distance.
Clumps which pass each other will also have bow shocks that exhibit changing intersection angles.
Thus, it is theoretically possible to observe Mach stems at the intersections of bow shocks for a wide range of clump separation distances.

\subsection{Instabilities}
\label{subsec:inst}
There are a number of instabilities that are relevant to these simulations.
A shocked clump will be compressed by a transmitted shock.
In the wake of the shock the clump will be accelerated and will exhibit Rayleigh-Taylor (RT) instabilities.
Shocked clump material will also expand laterally, interacting with post-bow shock gas leading to strong shear flows and Kelvin-Helmholtz (KH) instabilities.
Although we will not discuss this in great detail in this paper, we note that laboratory experiments have shown that cooling instabilities can play an important role in clump formation and fragmentation \citep{Suzuki-Vidal15}.

To consider the strength of these unstable modes we first recognize that the transmitted shock will cross the clump in a ``cloud-crushing'' time $t_{cc}$ \citep{Klein94},
\begin{equation}
    t_{cc} \simeq \frac{\sqrt{\chi}r_c}{v_s} ,\ \label{3}
\end{equation}
where $\chi$ is the clump-to-ambient density ratio, $r_c$ is the clump radius, and $v_s$ is the incident shock velocity.
The clump will be completely destroyed and mix with the post-shock flow within a few cloud-crushing times \citep{Yirak10}.

The RT instability will be most apparent at the head of the clump where dense material is being accelerated into lighter material.
For strong shocks, the approximate RT growth time is
\begin{equation}
    t_{RT} \simeq \frac{t_{cc}}{\sqrt{kr_c}} ,\ \label{4}
\end{equation}
where $k$ is the wave number of the instability.
According to equation \eqref{4}, the smallest wavelength modes will grow the fastest, but detailed simulation studies have shown that in the non-linear regime the dominant wavelength is of order the clump radius \citep{Jones96,Yirak10}.
Thus, the RT instability will grow on time scales of order the cloud-crushing time $t_{cc}$.

The KH instability will be seen on the edges of the clump where strong shear flows are generated.
For strong shocks, the KH growth time can be approximated as
\begin{equation}
    t_{KH} \simeq \frac{t_{cc}}{kr_c} .\ \label{5}
\end{equation}
Once again simulation studies show the most disruptive wavelength will be of order the clump radius \citep{Klein94,Poludnenko02}.
Thus, the relevant time scale for both RT and KH instabilities is of order the cloud-crushing time $t_{cc}$.

Cloud-crushing timescales were calculated for the relevant shock velocities in our simulations; the M = 7 clump has a $t_{cc} = 40.66$ years, the M = 10 clump has a $t_{cc} = 28.46$ years, and the M = 15 clump has a $t_{cc} = 18.98$ years.
Given a simulation time of 75 years, we therefore expect to see evolution of unstable modes into the nonlinear regime, (where instabilities show strong feedback on the flow), reached earlier in the two faster clumps compared to the M = 7 clump.
Note that as the clumps are destroyed, clump material will mix with the post-bow shock material, changing temperature and density conditions and, therefore, affecting the observed emission maps.

\begin{figure}
\includegraphics[width=0.5\textwidth]{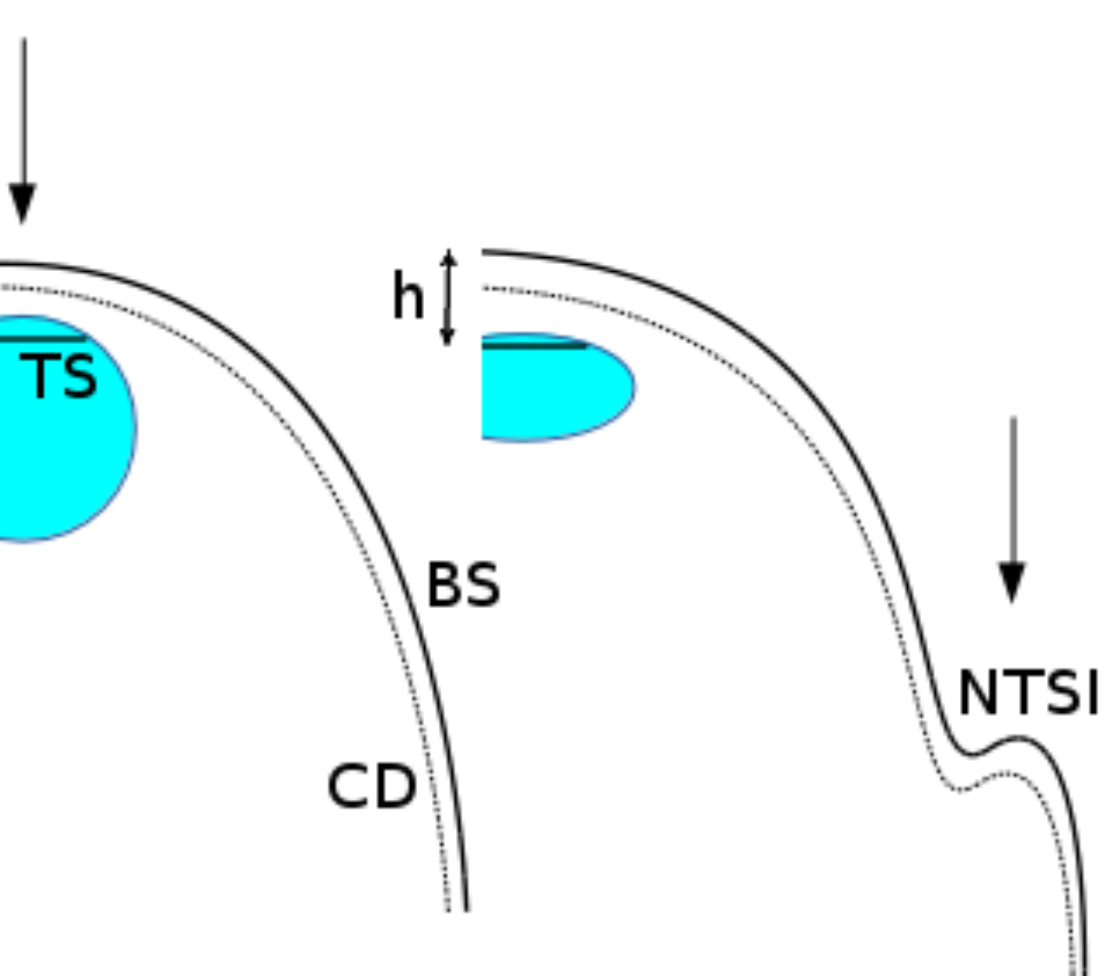}
\caption{Schematic showing shock structure and instabilities in a shocked clump and its bow shock.
The acronyms are as follows: ``BS'' = bow shock, ``CD'' = contact discontinuity, and ``TS'' = transmitted shock.
The large arrows represent the direction of the flow, and the double-arrow labeled ``h'' marks the thickness of the layer between the bow shock and transmitted shock.
The feature labeled ``NTSI'' represents a ripple From the Non-linear Thin Shell Instability that has been advected downstream.
One can imagine summing the emission along a line of sight perpendicular to this ripple will lead to an enhancement in H$\alpha$.}
\label{fig:inst}
\end{figure}

The emission will also be affected by any instabilities in the bow shocks themselves (as opposed to the clumps).
The bow shocks are susceptible to the aforementioned NTSI because strong cooling leads to a thin post-shock region bounded by the bow shock and the transmitted shock (see Figure~\ref{fig:inst}). 
Due to low thermal pressure in the cold material, this shock-bounded region will be dominated by the ram pressure of the incoming flow.
Any imbalance in the directions of the ram pressures on either shock can enhance perturbations of the shock interface and allow the NTSI to grow.
An NTSI-produced ripple at the head of the bow shock will be advected downstream as shown in Figure~\ref{fig:inst}. 
Thus the observed emission will appear more heterogeneous due to this rippling effect compared with emission from pre-existing clumps alone.
Furthermore, as the amplitude of the NTSI grows it can lead to fragmentation of the bow shock/clump complex if the clump itself is already beginning to lose its coherence.

The NTSI should have a fragmentation time of order the sound crossing time for the layer bounded by the two shocks (in this way it is similar to the Vishniac instability \citet{Vishniac83}). 
In our simulations, the thickness of this layer $h$ will be related to the cooling length $L_{cool}$.
Flows with stronger cooling will have shorter cooling lengths (i.e. smaller $h$) and thus be more easily disrupted by the instability and fragment on shorter time scales. 
The fragmentation time after the onset of the instability can be expressed as,

\begin{equation}
    t_{f} \simeq \frac{h}{c_s} ,\ \label{6}
\end{equation}
where $c_s$ is the sound speed and $h$ is the thickness of the bounded layer.

We ran 1-D radiative shock models in order to determine the cooling lengths in our simulations. 
We found a cooling length $L_{cool}$ of approximately 0.08 clump radii for the head of the bow shock of the M = 15 clump, and approximately 1.27 $r_c$ for the M = 7 clump.
With an $L_{cool}$ that is approximately 15.6 times shorter, we expect the bow shock of the M = 15 clump to fragment much faster than the bow of the M = 7 clump once the instability has begun to disrupt the bows.
This difference should also be apparent in the emission maps as more fragmentation will lead to a larger heterogeneous region which we call the froth.

\section{Results}
\label{sec:results}
The evolution of structure in our simulations is driven by clumps, their bow shock interactions and the growth of instabilities.
Bow shock intersections will lead to either regular or Mach reflection as discussed in Section~\ref{subsec:machstem}, and the emission will be enhanced at these intersection points.
As discussed in Section~\ref{subsec:inst}, the clumps will be susceptible to the Rayleigh-Taylor and the Kelvin-Helmholtz instabilities on a time scale of order the cloud-crushing time $t_{cc}$ \eqref{3}.
Due to the presence of strong cooling, the bow shocks are unstable to the NTSI and will fragment leading to enhanced heterogeneous structuring.
We see signatures of these intersections and instabilities in the column density and synthetic emission maps that follow.
In the following subsections, we refer to the models by their letter designation (see  Table~\ref{tab:sims}).

\subsection{Bow Shock Interactions}
\label{subsec:intersection}
\begin{figure*}
\centering
\includegraphics[width=\textwidth]{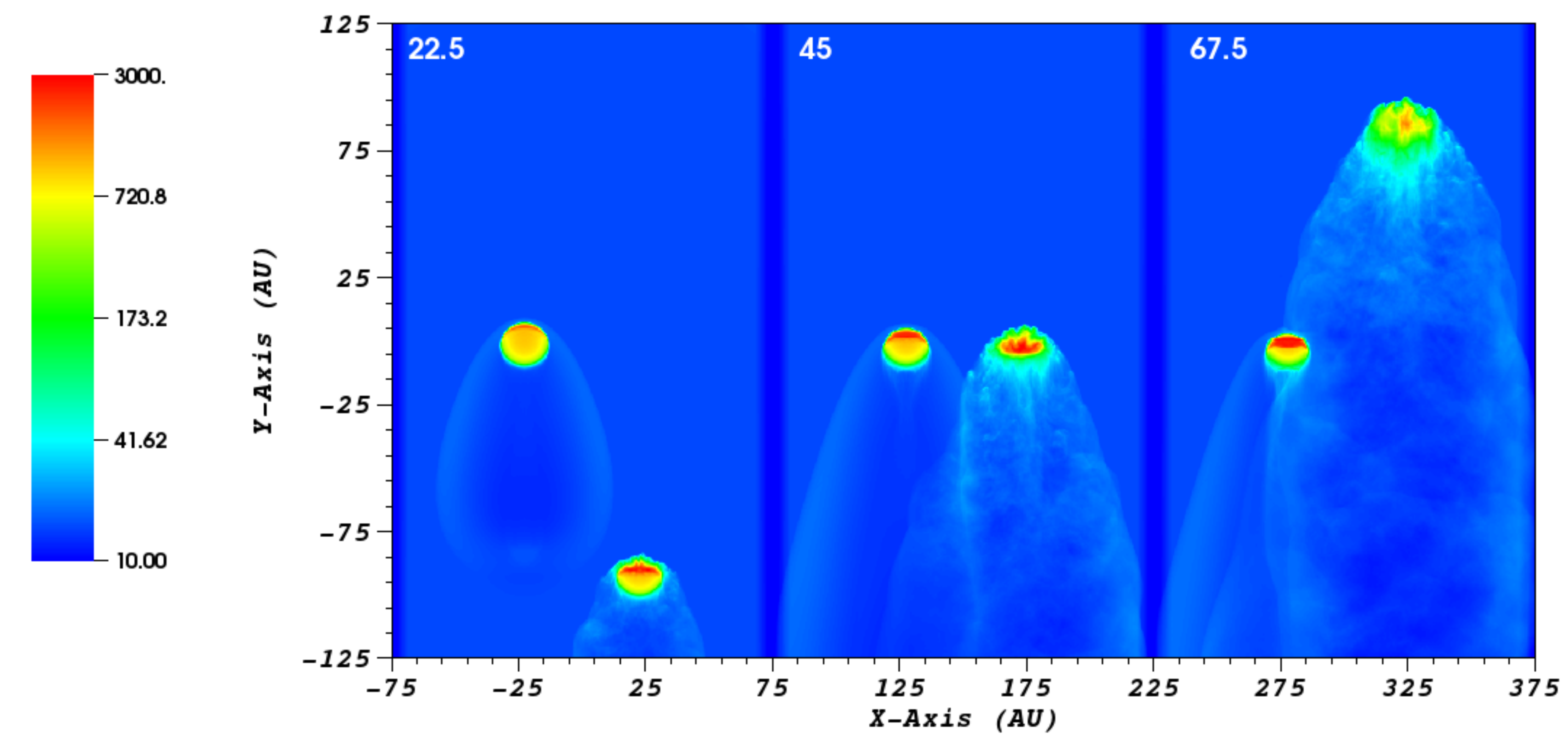}
\caption{Morphology of two intersecting, co-moving bow shocks and their parent clumps.
Shown is a time series, from run E, of column density in units of 1000 cm\textsuperscript{-3}.
Panels are labeled left to right by their simulation time in years (22.5, 45, 67.5 respectively).
The intersection point of the two bow shocks is clearly visible in the middle panel and moves laterally to the left.}
\label{fig:Edens}
\end{figure*}

We first consider the 2-clump models (runs A - F).
Figure~\ref{fig:Edens} shows a column density map of run E at simulation times of 22.5, 45, and 67.5 years.
In this figure, we can clearly see the intersection point of the two bow shocks in the middle panel.
Given the time dependent nature of the flows it is difficult to ascertain the presence of Mach stems, however we can draw conclusions about the nature of the shock interactions.
In regular reflection, gas will travel through two shocks: the incident bow and the reflected shock (see Figure~\ref{fig:reflect}).
In Mach reflection, the gas will travel through a strong, planar shock.
Hence, in either scenario, gas will experience a higher density compression and temperature jump than it would from a single bow shock, and this will lead to enhanced emission at the intersection point as is seen in the figures.

\begin{figure*}
\includegraphics[width=\textwidth]{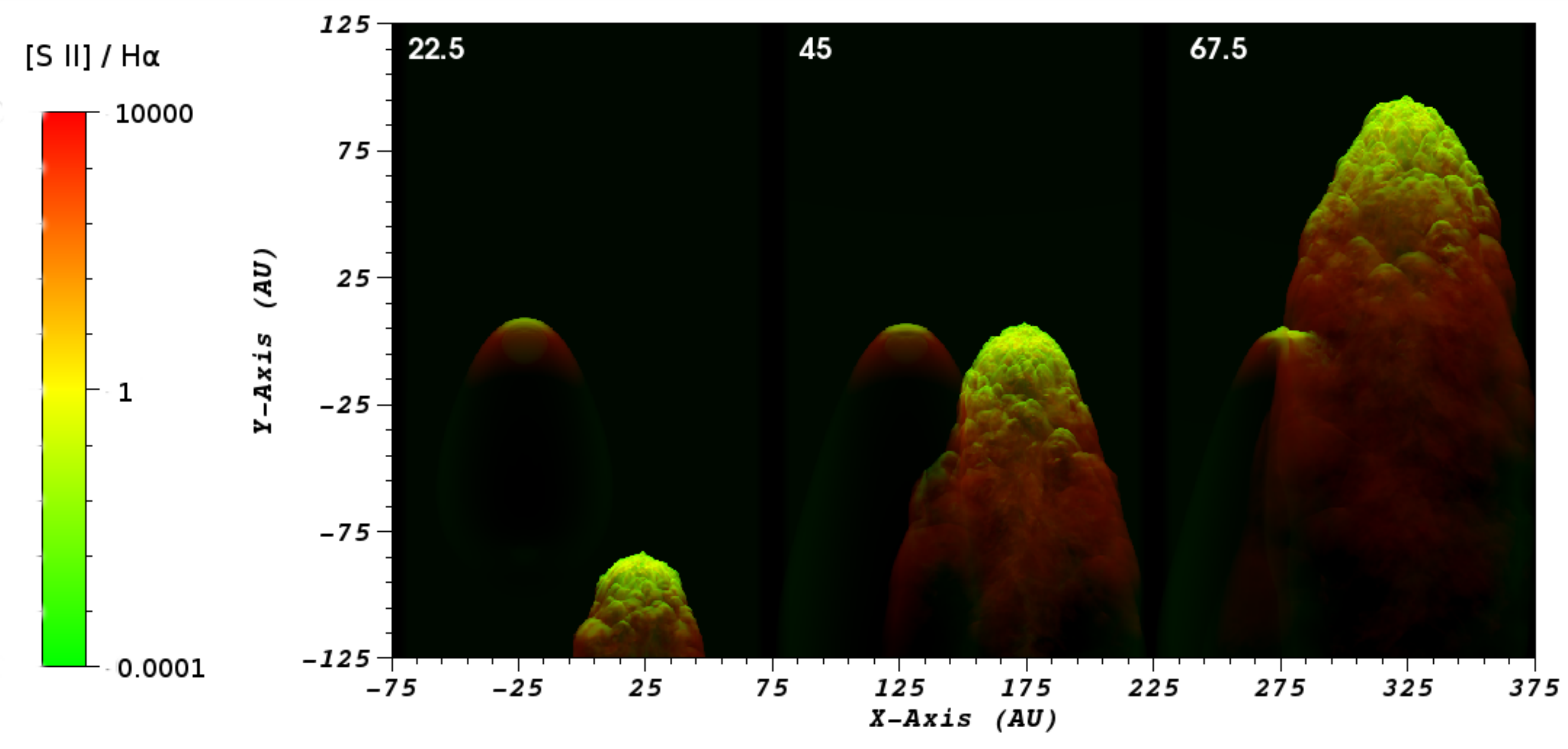}
\caption{Time evolution of emission line structure of two intersecting, co-moving bow shocks.
The scaling of the ratio of [S II] to H$\alpha$ is shown, resulting in green for H$\alpha$ dominated regions, red for [S II] dominated regions, and yellow in areas where both lines are strong.
Panels are from same model (E) and at the same simulation times as Figure~\ref{fig:Edens}.
Note how the emission is brighter at the intersection point in the middle panel relative to the neighboring bow shocks.}
\label{fig:Eemiss}
\end{figure*}

We now turn to the emission for run E which is shown in Figure~\ref{fig:Eemiss} at the same simulation times as in Figure~\ref{fig:Edens}.
H$\alpha$ marks shock fronts in green, [S II] marks cooling regions in red, and yellow appears in regions where there is strong emission from both lines.
The bow shocks are brightest in H$\alpha$ at the leading edges where the shocks are less oblique.
However, the intersection point of the two bow shocks where Mach stem formation is possible will also be bright in H$\alpha$.

\begin{figure*}
\includegraphics[width=\textwidth]{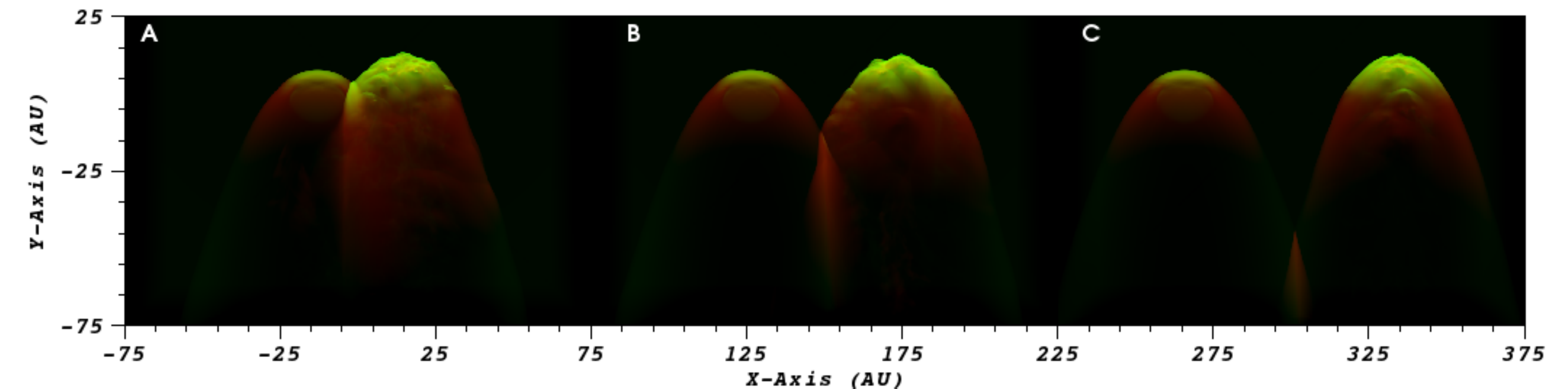}
\caption{Effect of clump separation on emission line structure.
Panels from left to right are from runs A, B, and C respectively at a simulation time of 42 years.
Scaling is the same as in previous emission map figures.
Clump separation affects the location of the bow shock intersection point as well as the intensity of the emission at said point.}
\label{fig:ABCemiss}
\end{figure*}

Figure~\ref{fig:ABCemiss} shows how clump separation affects the emission of intersection points.
Recall how increasing separation of the clumps translates into a smaller intersection angle $\alpha$.
We see the bow shock intersection point in all 3 models (A, B, and C), but the location and intensity of the emission at the point varies.
When the clumps are farther apart, the bows intersect farther downstream at a steeper angle (i.e. $\alpha$ in Figure~\ref{fig:reflect} is smaller).
Mach stem formation requires a large $\alpha$ when strong cooling is present as discussed in Section~\ref{subsec:machstem}.
However, with an unstable corrugated bow shock, $\alpha$ may momentarily reach the critical value $\alpha_c$ even when the clumps are far apart.
Regardless of the presence of a Mach stem, we see that as clump separation decreases, the H$\alpha$ intensity of the intersection point increases.

\begin{figure}
\includegraphics[width=0.5\textwidth]{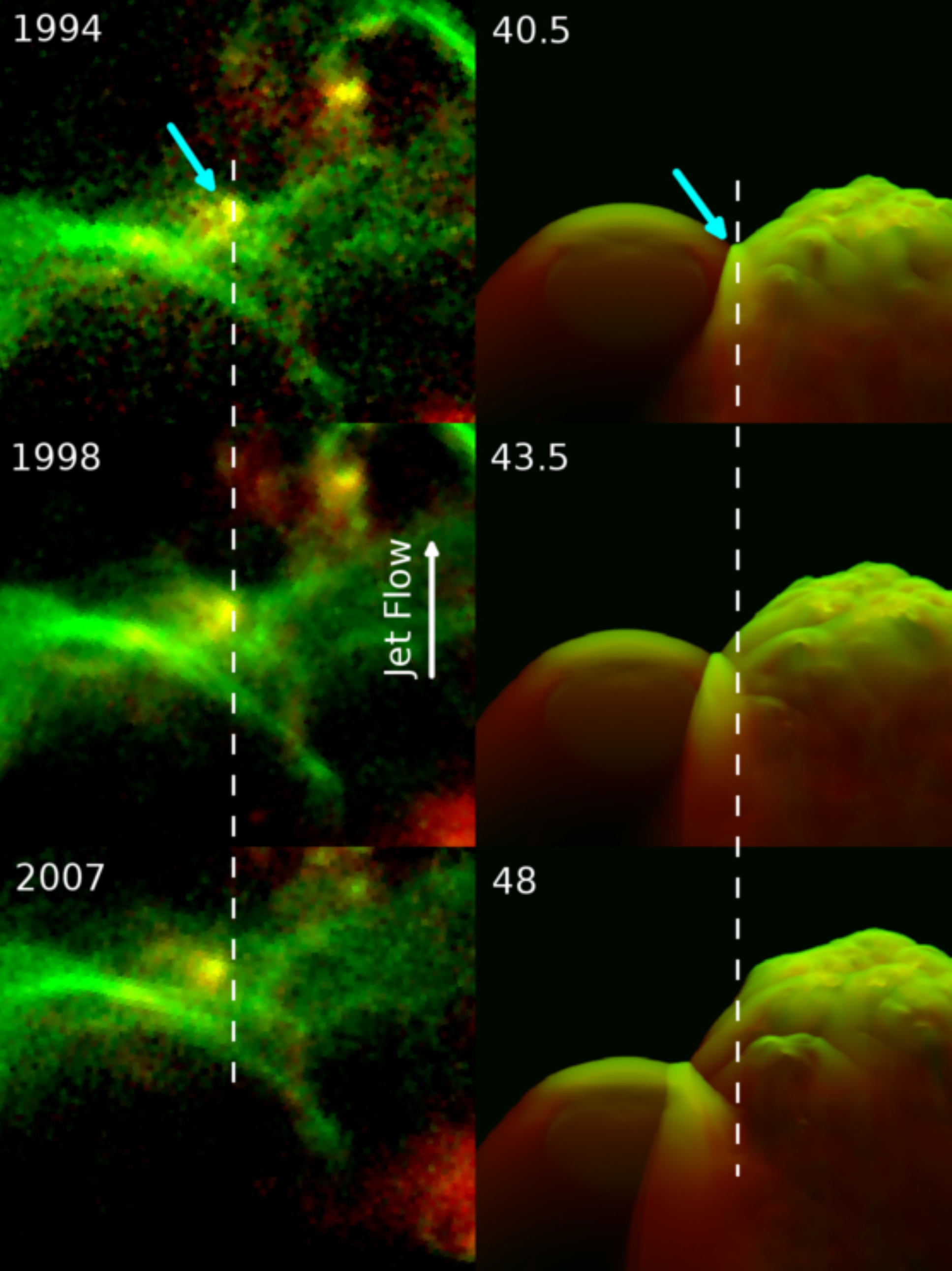}
\caption{Comparsion of HH 34 and run A.
The knot marked with the \emph{cyan} arrow in HH 34 (\emph{left}) exhibits lateral motion just as the bow shock intersection point (also marked with a \emph{cyan} arrow) in simulation A (\emph{right}).
The stationary dashed lines are included to see the displacements more easily.
The images of HH 34 are from the Hubble Space Telescope from 1994, 1998, and 2007 respectively from top to bottom, and these frame from simulation A were taken at simulation times of 40.5, 43.5, and 48 years.
All images show the bulk flow directed upwards, and H$\alpha$ is in green and [S II] is in red.
}
\label{fig:HH34}
\end{figure}

Consideration of figure~\ref{fig:Eemiss}, shows that by 67.5 years the intersection point has moved to the left and is now disrupting the head of the M = 7 clump/bow shock.
This type of lateral motion is seen in all of our simulations, and it is a signature of an intersection point which has been observed in objects such as HH 34 \citep{Hartigan11}.
Figure~\ref{fig:HH34} shows three images of HH 34 alongside three images from simulation A.
The knots marked with the \emph{cyan} arrows move laterally to the left compared to the bulk flow which moves directly upwards.

\begin{figure}
\includegraphics[width=0.5\textwidth]{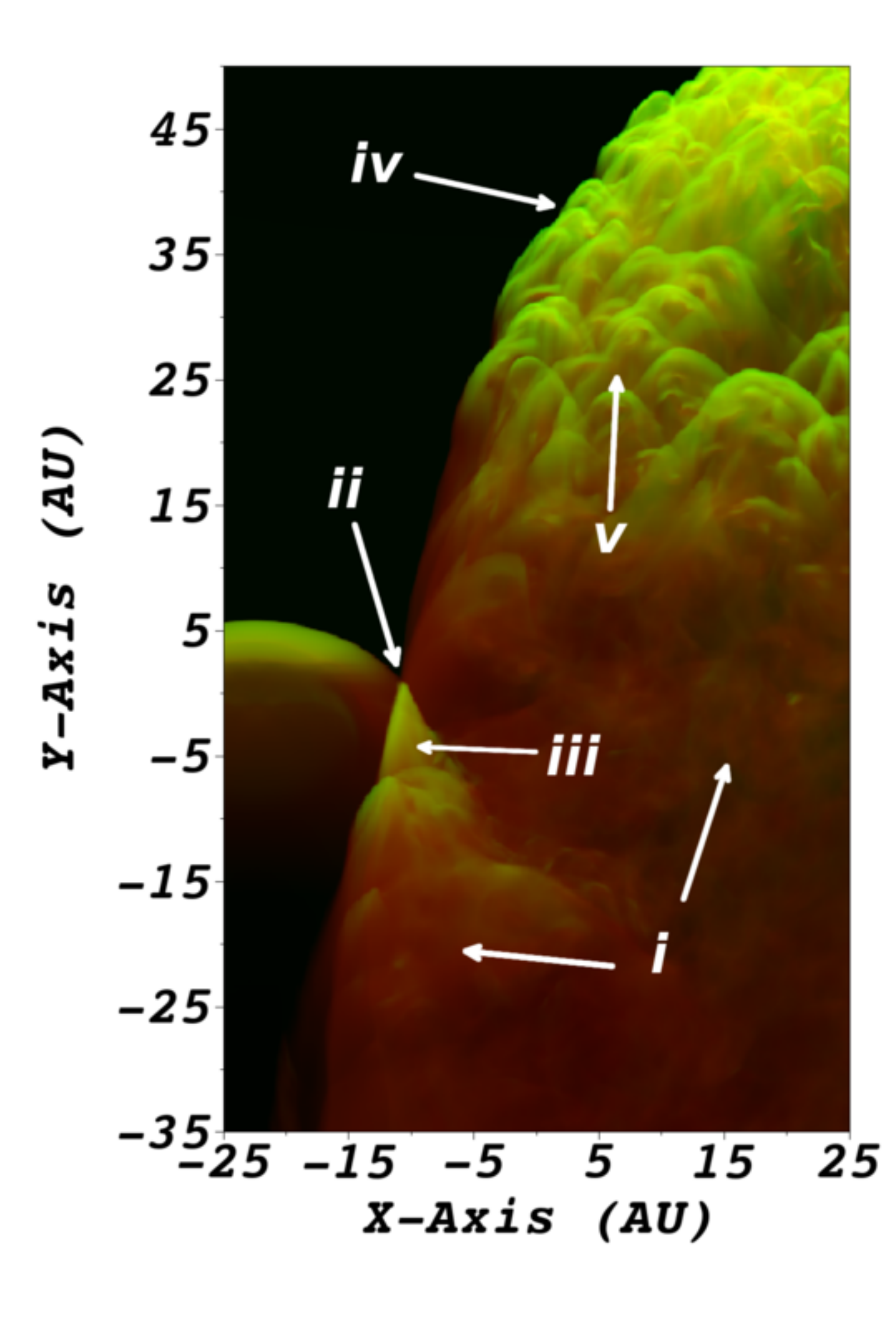}
\caption{Enlarged emission map showing details of intersecting bow shocks.
Image is from run E at a simulation time of 57 years, and scaling is the same as in Figure~\ref{fig:Eemiss}.
The arrows label the following features: $i = $ post-shock cooling regions, $ii = $ bow shock intersection point, $iii = $ post-intersection point ``wedge'', $iv = $ unstable bow shock, $v = $ ``froth''.}
\label{fig:Eemissenlarged}
\end{figure}

Figure~\ref{fig:Eemissenlarged} shows an enlarged emission map of run E at a simulation time of 57 years, and exemplifies several important structures that are generic to our study.
The arrows labeled $i$ mark the reddened post-shock cooling regions where [S II] is dominant.
These regions always appear behind the shocks marked by green H$\alpha$ emission.
The point $ii$ marks the intersection of the two bow shocks, and it is much brighter in H$\alpha$ than the oblique shocks to either side of it.
It will continue to move laterally to the left, and eventually disrupt the M = 7 clump/bow, as the faster bow moves forward.

Another noteworthy feature is the post-intersection ``wedge'' which is visible in Figure~\ref{fig:Eemissenlarged} labeled by point $iii$. 
Like the intersection point, this feature also moves laterally and will eventually blend in with the head of the M = 7 bow.
As mentioned in Section~\ref{subsec:machstem}, this wedge exists regardless of the presence of a Mach stem and is present when regular reflection occurs as well.
In HST images, these types of features are unlikely to appear so cleanly (i.e with such symmetry), but enhanced emission features are observed behind the intersection points of bow shocks \citep{Hartigan11}.

\subsection{Instabilities and Emission ``Froth''}
\label{subsec:froth}
The other two points in Figure~\ref{fig:Eemissenlarged} (points $iv$ and $v$) are related to the NTSI and emission heterogeneity or ``froth'' which were described in Section~\ref{subsec:inst}. 
Point $iv$ marks the edge of the unstable bow shock and point $v$ denotes the region of highly heterogeneous, patchy emission denoted by the term froth.
The corrugation and filamentary nature of the edge of the bow shock confirms that it is unstable.
This instability (the NTSI) produces the observed froth which we discuss further in this subsection.

\begin{figure}
\includegraphics[width=0.5\textwidth]{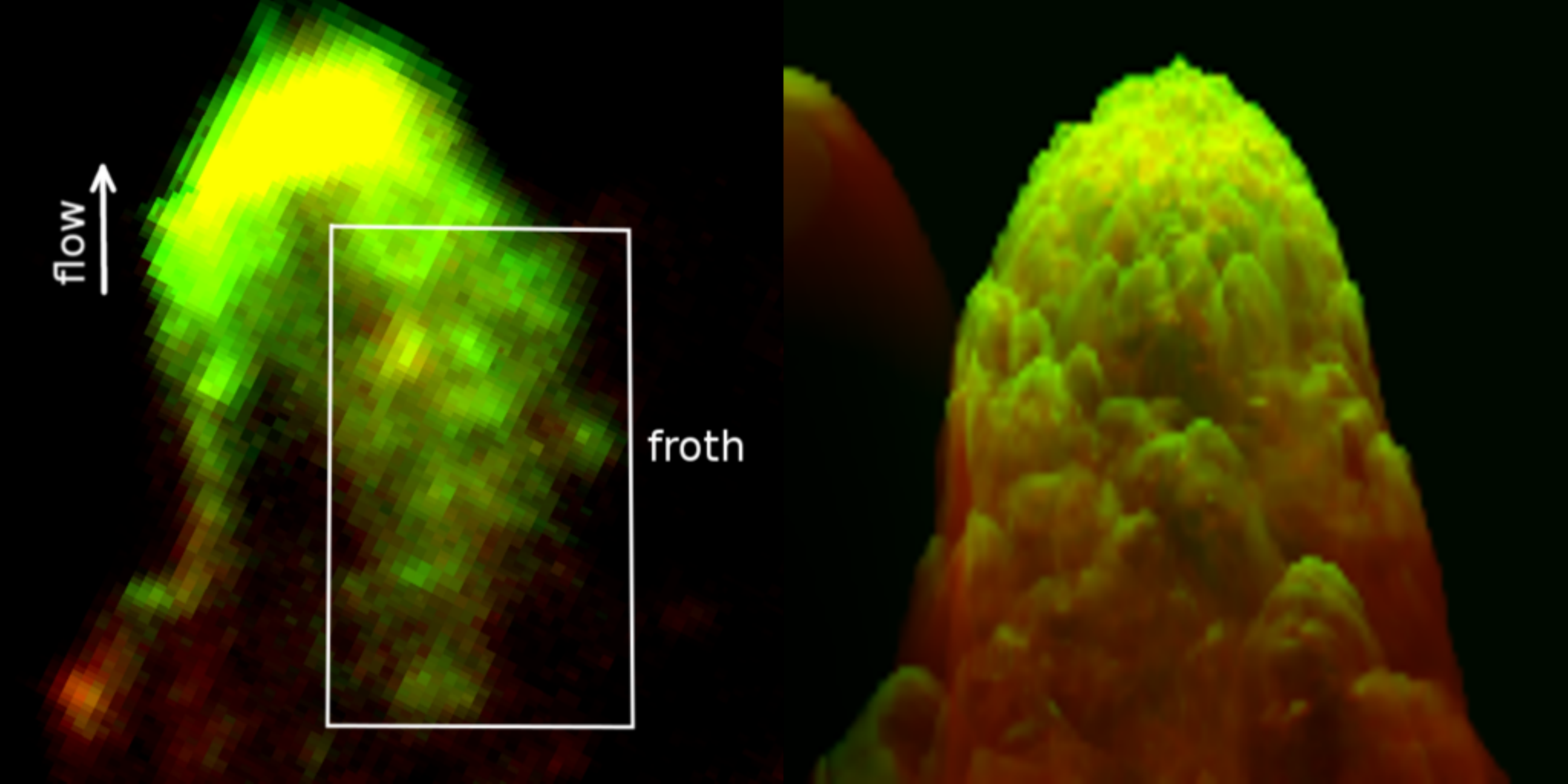}
\caption{Comparsion of HH 1 and run E.
HH 1 (\emph{left}) exhibits a post-shock region of froth just as in simulation E (\emph{right}).
The image of HH 1 is from the Hubble Space Telescope from 2007, and this frame from simulation E was taken at a simulation time of 45 years.
Both images show the flow directed upwards, and H$\alpha$ is in green and [S II] is in red.
}
\label{fig:HH1}
\end{figure}

Looking back at Figure~\ref{fig:Eemiss}, it is clear that the M = 15 bow shock has much more heterogeneous behavior in its wake as compared to the M = 7 bow shock.
Some of this may be due to clump material being stripped off by KH instabilities, but the main reason for this is the corrugation of the bow shock by the NTSI.
The instability causes the bow shock to fragment which leads to the observed filamentary structures.
In \citet{Hartigan11}, the same features were observed in HH objects such as HH 1, and they called this filamentary, heterogeneous region the ``froth''.
They suggested that the structures in the froth could have arisen from irregularities in the shape of the bow shock, and this appears to be the case in our simulations.
Figure~\ref{fig:HH1} shows a direct comparison of HH 1 with run E.
Note that in both observation and simulation, we see the heterogeneity in the emission of both real and simulated flows.

We have run separate simulations to confirm  when cooling is turned off we do not see the corrugation of the shocks associated with the NTSI.  
Note that the NTSI is also visible in the column density maps of our simulations (see e.g. Figure~\ref{fig:Edens}).
The bow shocks from both clumps are clearly visible, and they are clearly not smooth paraboloids.
The effect of the NTSI is more pronounced in the M = 15 bow because the cooling length is shorter than that of the M = 7 bow (stronger shock implies shorter cooling length).
This in turn implies a shorter distance between the bow shock and the contact discontinuity $h$ which will reduce the growth time of unstable modes $t_f$ \eqref{6}.

Also visible in the column density maps are the transmitted shocks which propagate through and compress the clumps.
The transmitted shock is unable to fully disrupt the M = 7 clump during the simulation confirming that it takes a few cloud-crushing times to destroy clumps \citep{Yirak10} ($t_{cc} = $ 40.66 years for this clump).
The M = 15 clump, on the other hand, has a cloud-crushing time of only 18.98 years and thus the transmitted shock can fully compress and disrupt the clump within the simulation time.
This faster clump also exhibits features that are characteristic of the RT instability; RT spikes are visible at the head of the clump as early as 22.5 years, and RT bubbles penetrate the clump. 
There is also evidence for KH instabilities as material is stripped away from the lateral edges of the clump.
The ratio of equation \eqref{4} to equation \eqref{5} demonstrates that $t_{RT} > t_{KH}$ for modes smaller than the clump radius $r_c$.
From this we might conclude that the clump destruction will be dominated by the KH modes.
The conclusion, however, would be based only on order of magnitude estimates from linear theory.
The simulation results suggest that it is the RT instability that dominates in driving the break-up of the clump.

\begin{figure*}
\includegraphics[width=\textwidth]{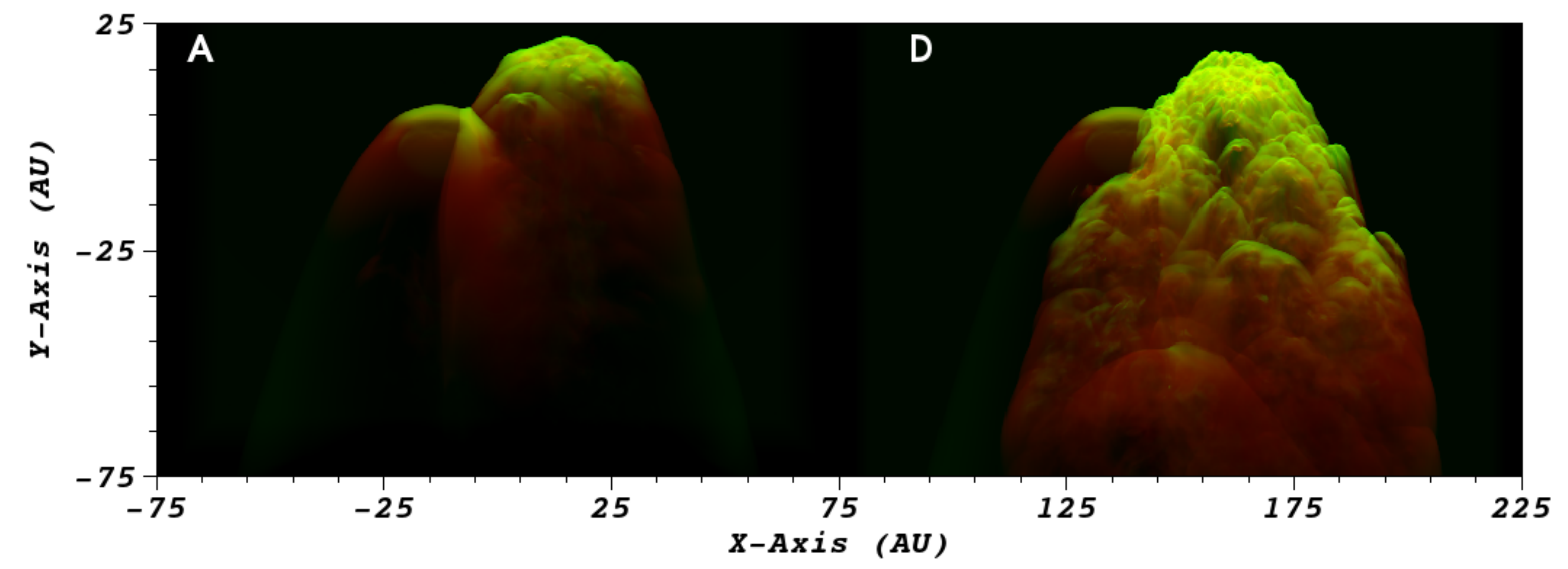}
\caption{Effect of clump and bow shock velocity on emission line structure.
Left panel is from run A and right panel is from run D at a simulation time of 48 years.
Scaling is the same as in previous emission map figures.
Higher velocities lead to more unstable behavior in the clump and bow shock which is marked by stronger, disorganized emission.}
\label{fig:ADemiss}
\end{figure*}

The time scales of all of the instabilities are dependent on the shock velocity.
Thus, a side-by-side comparison of models differing only by velocity is useful.
Figure~\ref{fig:ADemiss} shows a comparison of runs A and D; model A has a M = 10 clump and model D has a M = 15 clump.
In run A, we can clearly see the intersection point and the wedge while in run D the instabilities dominate the region and make it difficult to observe the intersection point.
The stronger H$\alpha$ in run D versus run A can be explained based on 3 physical attributes of the runs.
First, a stronger shock in run D implies higher post-shock densities and temperatures which leads to stronger H$\alpha$ emission.
Secondly, the M = 15 clump has a shorter cloud-crushing time $t_{cc}$ than the M = 10 clump (see Section~\ref{subsec:inst}), hence run D is more susceptible to KH instabilities that can strip off clump material leading to a more heterogeneous post-shock region.
Finally, the stronger shock has a shorter cooling length and hence a smaller layer thickness $h$ and shorter fragmentation time $t_f$ (see equation \eqref{6}).
Thus, the M = 15 bow is more susceptible to fragmentation via the NTSI which explains why run D has a larger region of froth as compared to run A.

\subsection{Clump Orientation and Viewing Angle}
\label{subsec:viewingangle}
In order to understand the role of clump orientation and viewing angle we now consider the 3-clump runs.
Recall that Figures~\ref{fig:3-clumptop} and~\ref{fig:3-clumpside} show the geometry applied in these models, and the reader should refer to Table~\ref{tab:sims} for the defining simulation parameters that are varied from model to model.
Structures observed in the synthetic emission images change or even become invisible depending on the relative orientation of the clumps or the observer's viewing angle.
We note that laboratory experiments have demonstrated how clump orientation affects the appearance of bow shocks \citep{Hartigan09}.
Likewise, relative observational viewing angle determines observational appearances.
For example, a bow shock moving in the plane of the sky appears more collimated than if it were directed towards the observer.

\begin{figure*}
\includegraphics[width=\textwidth]{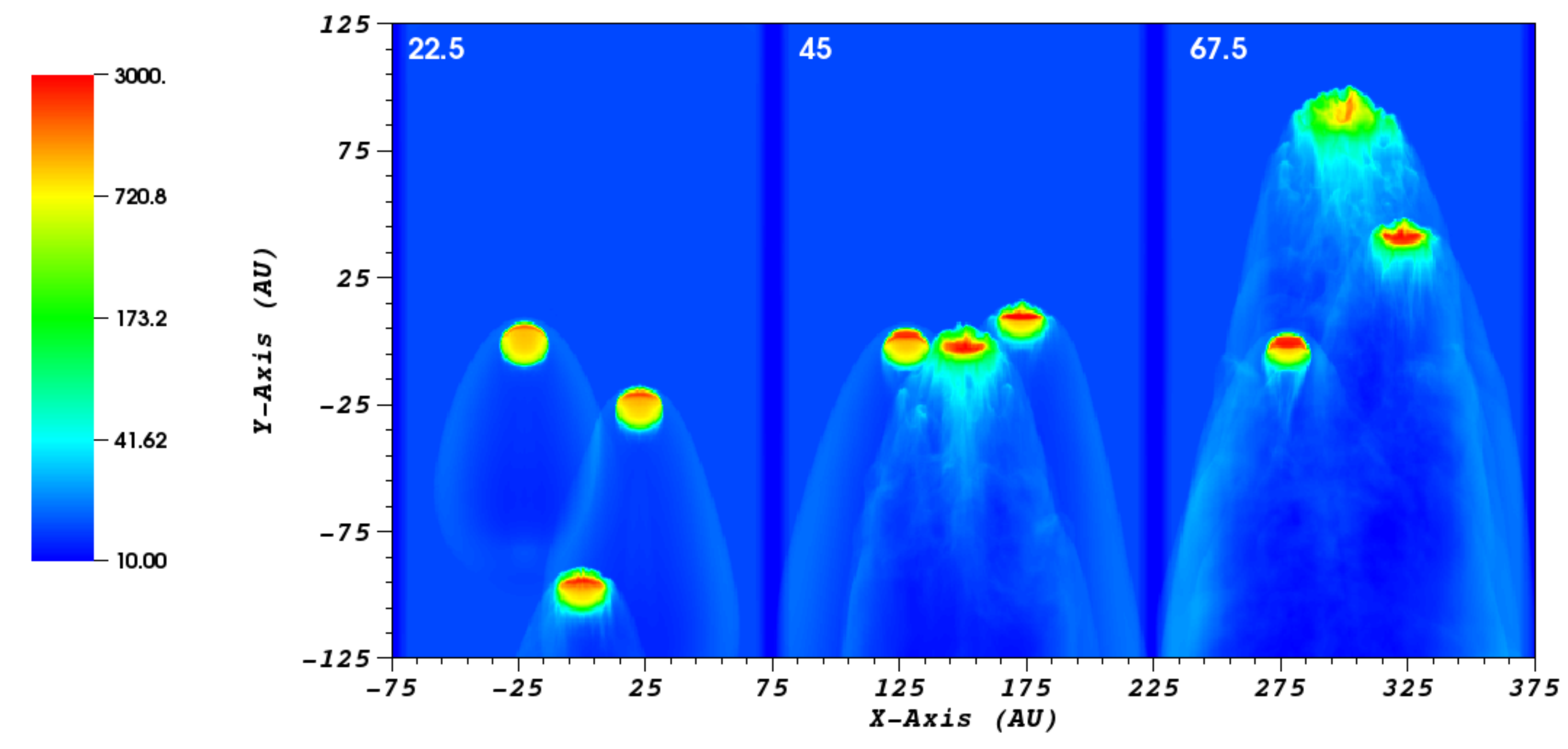}
\caption{Morphology of three intersecting, co-moving bow shocks and their parent clumps.
Shown is a time series, from run K, of column density in units of 1000 cm\textsuperscript{-3}.
Panels are labeled left to right by their simulation time in years (22.5, 45, 67.5 respectively).
The evolution of the fastest moving clump is now altered by the third clump's bow shock (compare with Figure~\ref{fig:Edens}).}
\label{fig:Kdens}
\end{figure*}

Figure~\ref{fig:Kdens} shows a column density map time series for run K.
We immediately observe that introducing a third clump introduces new complexity into the flow.
Each clump and bow shock is now affected by two other bow shocks simultaneously.
Furthermore, there are now 3 bow shock intersection points.
Note that depending on orientation, these intersections may not always be visible.
For example, the intersection point between the M = 7 and M = 10 bows is visible at 22.5 and 67.5 years, but not at 45 years due to the obstructing M = 15 clump.

\begin{figure*}
\includegraphics[width=\textwidth]{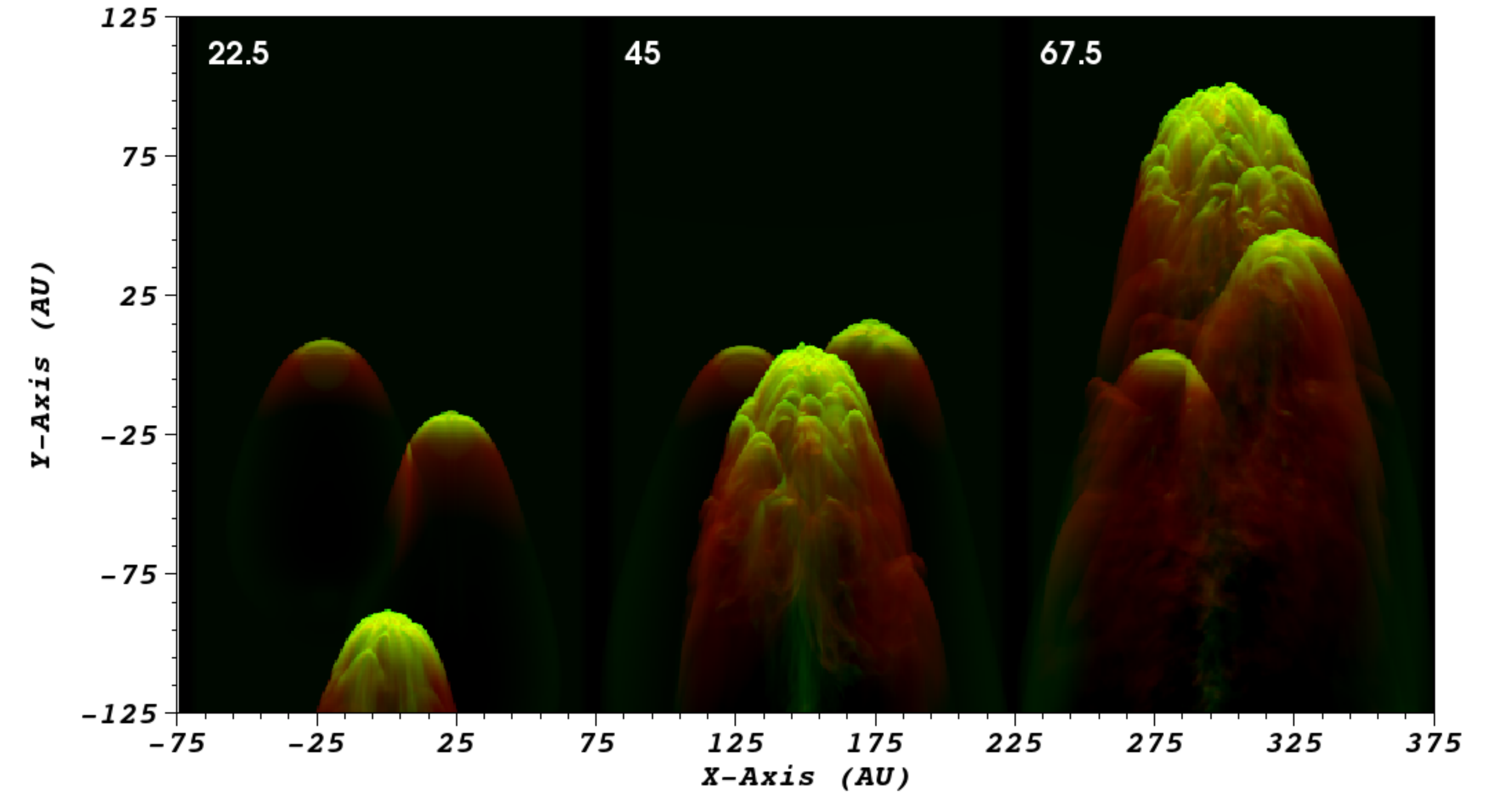}
\caption{Time evolution of emission line structure from three intersecting, co-moving bow shocks.
Panels are from same model (K) and at same simulation times as Figure~\ref{fig:Kdens}.
Scaling is the same as in previous emission map figures.
At times, certain features, such as bow shock intersection points, can become impossible to see due to the presence of an obstructing, third clump.}
\label{fig:Kemiss}
\end{figure*}

Figure~\ref{fig:Kemiss} shows the corresponding time series of emission for run K.
The features and time-dependent evolution discussed in the previous subsections (the 2-clump runs) are observed here as well.
At 22.5 and 67.5 years  an intersection point is apparent which appears as a bright spot of H$\alpha$ emission.
Note that this point has moved laterally to the left between these two frames and is accompanied by a post-intersection point wedge as well.
As the M = 15 clump nears the other clumps, its emission dominates the region and interactions between the other clumps are not visible.
As we have seen in previous simulations, the fastest moving clump is most susceptible to instabilities and thus exhibits a larger region of froth.

\begin{figure*}
\includegraphics[width=\textwidth]{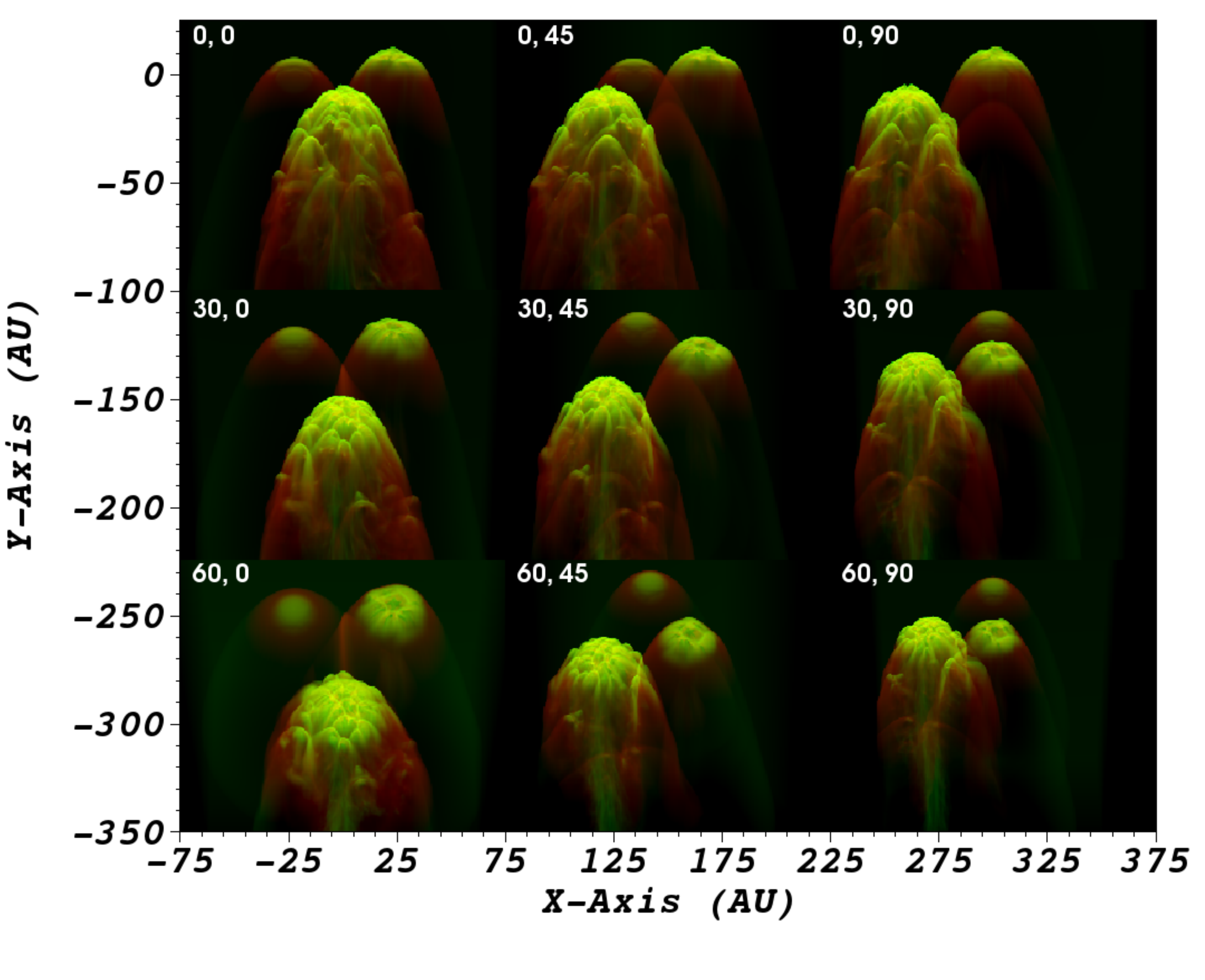}
\caption{Effect of the observer's viewing angle on emission line structure.
Panels are labeled by their inclination angle $i$ and rotation angle $\upvarphi$ respectively (see Figures~\ref{fig:3-clumpside} and~\ref{fig:3-clumptop}).
All panels are from run K at a simulation time of 42 years, and scaling is the same as in previous emission map figures.
Viewing angle can drastically change the observed location as well as the emission intensity of certain features such as bow shock intersection points.}
\label{fig:Kemissangles}
\end{figure*}

We now consider a single frame from run K and change the angle of the observer to understand how altering viewing orientation affects the synthetic emission maps.
Figure~\ref{fig:Kemissangles} shows 9 different angles: 3 inclination angles and 3 rotation angles.
Increasing the inclination angle points the direction of the outflow towards the observer, and the rotation angle rotates the viewpoint around the axis of the direction of flow (see Figures~\ref{fig:3-clumptop} and~\ref{fig:3-clumpside}).
As the inclination angle is increased, the emission from the bow shocks appears more circular.
This occurs as the bow shocks' paraboloid projection on the plane of the sky begins to catch both the front and back side of the bow..

The change in the appearance of the intersection point with inclination angle is also noteworthy.
The intersection of the bow shocks appears more like a line at high inclination angles, and the emission appears brighter as well.
The increase in surface brightness occurs as the line of sight along which the emission is summed now passes through more of the post-intersection point wedge.

Changing the rotation angle has the effect of rendering flow features visible to a greater or lesser degree.
At some angles, the intersection between the M = 15 and M = 10 bows are apparent while the intersection between the M = 10 and M = 7 bows are invisible.
Furthermore, in the $i = 0$, $\upvarphi = 90$ panel, the M = 7 clump is completely obstructed by brighter flow features making this emission map appear as if there are only 2 clumps.
Changing the relative positions of the clumps has a similar effect.

\begin{figure*}
\includegraphics[width=\textwidth]{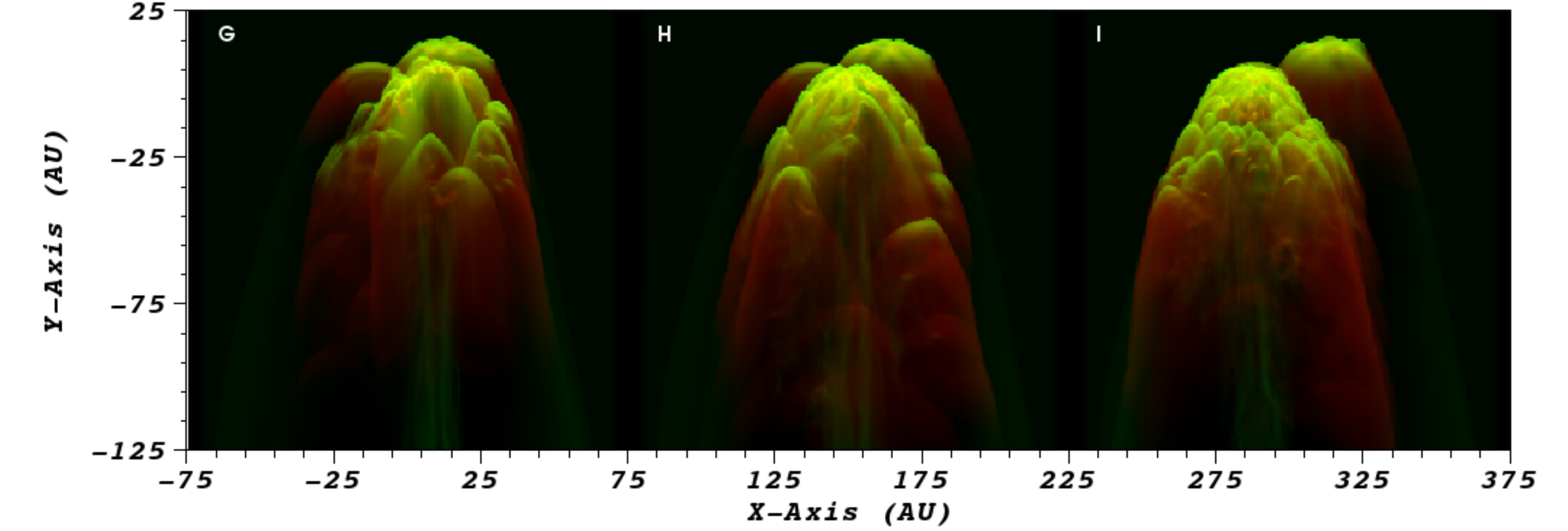}
\caption{Effect of clump orientation on emission line structure.
Panels from left to right are from runs G, H, and I respectively at a simulation time of 45 years.
Scaling is the same as in previous emission map figures.
Varying the position of a third clump can obstruct or make visible features such as bow shock intersection points.}
\label{fig:GHIemiss}
\end{figure*}

Figure~\ref{fig:GHIemiss} shows how changing the relative position of the third, M = 15 clump can affect the emission.
As Table~\ref{tab:sims} shows, runs G, H, and I differ only by their orientation angle $a$ (Figure~\ref{fig:3-clumptop} ).
The effect is similar to changing the viewing angle, but the cause is different.
For example, the intersection of the M = 10 and M = 7 bows is only visible in the run G panel, and only two clumps and bow shocks are visible in the run I panel.

\subsection{``Shock Curtains'': The H$\alpha$ Sheet}
\label{subsec:sheet}
\begin{figure*}
\includegraphics[width=\textwidth]{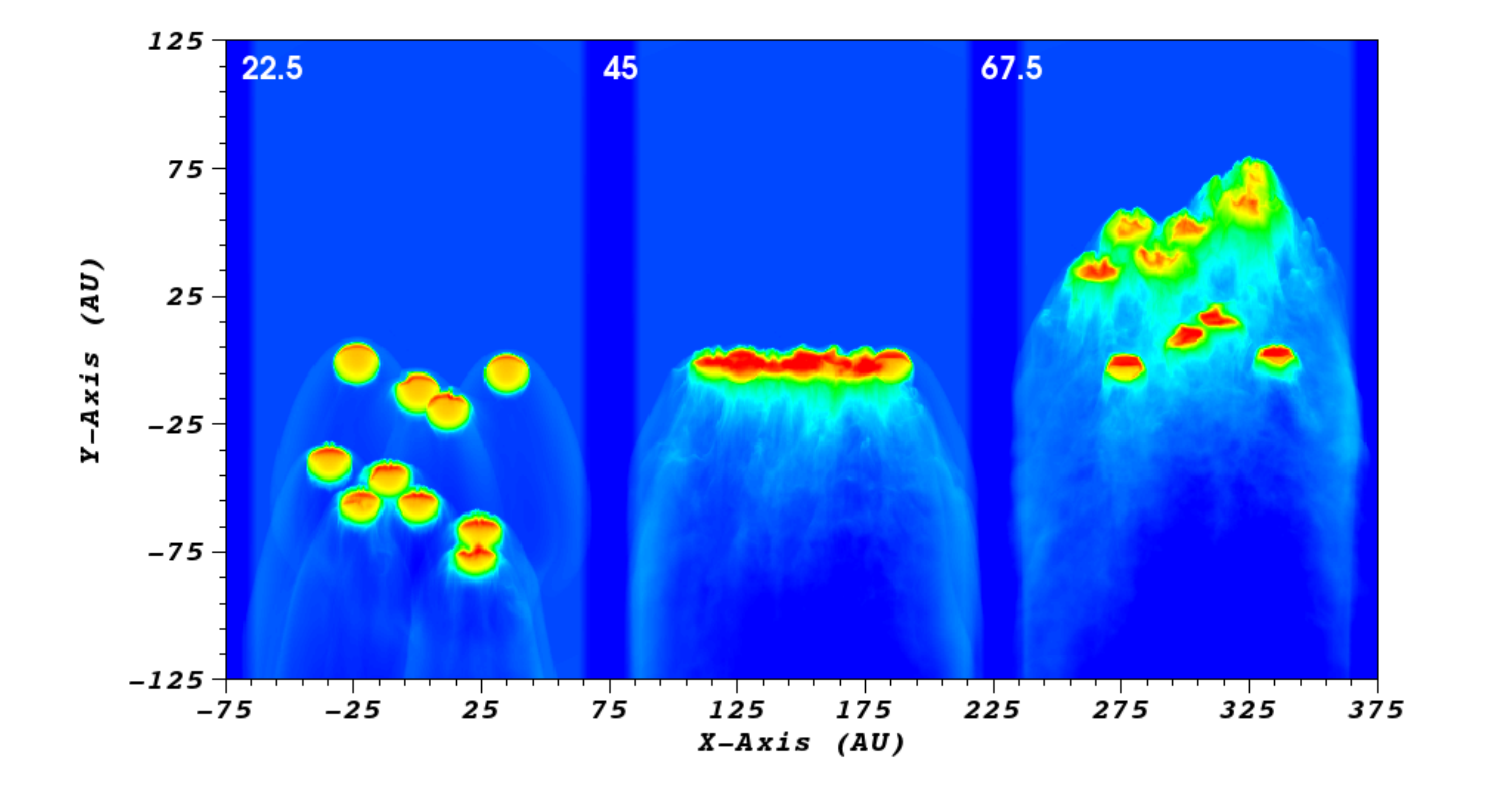}
\caption{Intersecting, co-moving bow shocks from multiple parent clumps with random velocities. 
Figure shows a time series of column density from run M in units of 1000 cm\textsuperscript{-3}.
Panels from left to right are at simulation times of 22.5, 45, and 67.5 years respectively.
In the middle panel, the clumps cross the same plane at approximately the same time forming a sheet of shocked material.}
\label{fig:Mdens}
\end{figure*}

\begin{figure*}
\includegraphics[width=\textwidth]{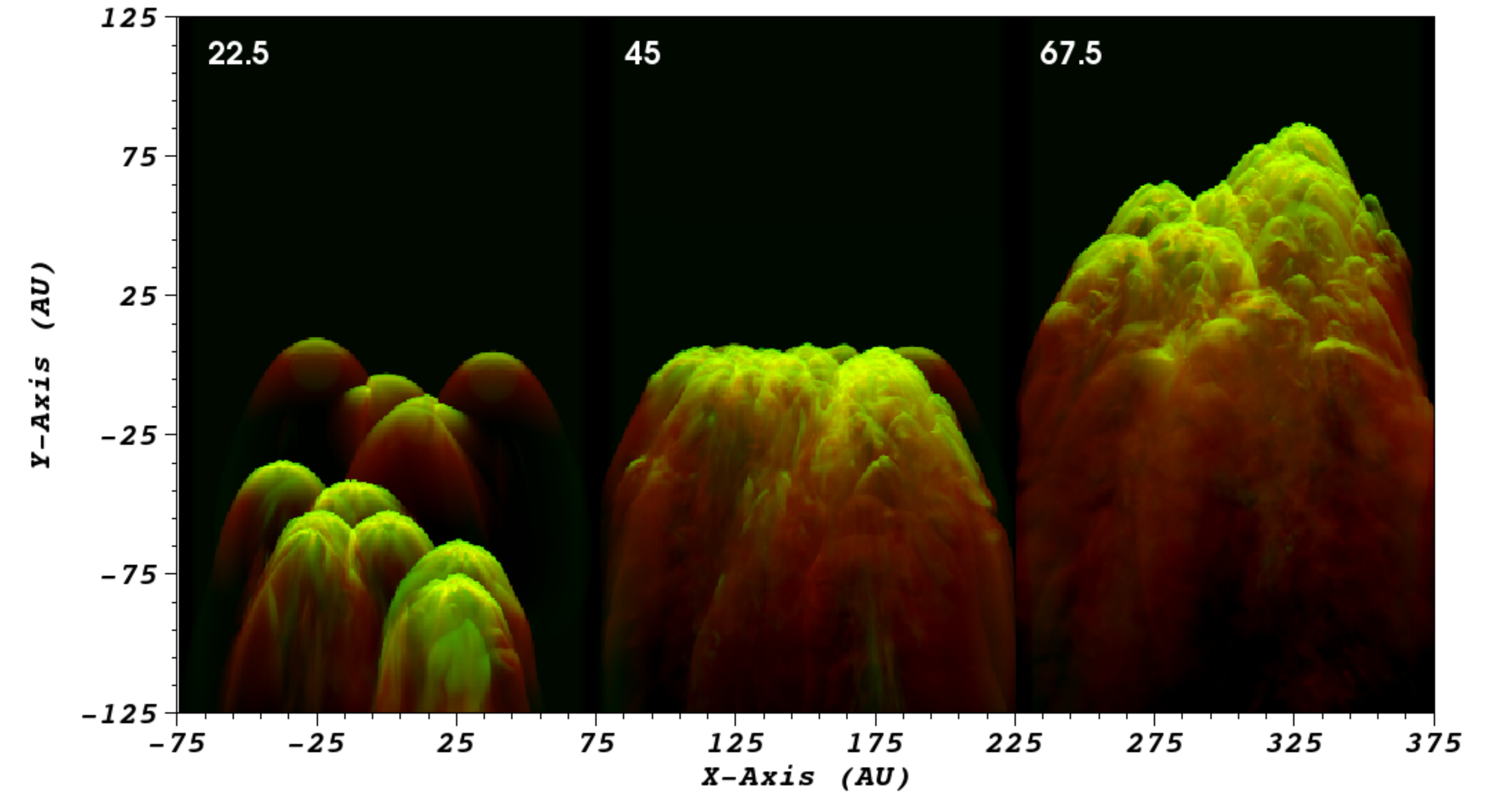}
\caption{Time evolution of emission line structure from multiple bow shocks with random velocities.
Panels from left to right are from run M at simulation times of 22.5, 45, and 67.5 years respectively.
Scaling is the same as in previous emission map figures.
The multiple clumps and bow shocks form a ``sheet'' of H$\alpha$ emission accompanied by complex emission structure in the post-shock regions.}
\label{fig:Memiss}
\end{figure*}

We now consider the multi-clump model.
Figure~\ref{fig:Mdens} shows a time series of column density maps from run M, and Figure~\ref{fig:Memiss} shows the corresponding time series of emission.
Given the number of clumps involved it is not surprising that we see a highly structured and complex flow.
The first panel of Figure~\ref{fig:Mdens} illustrates the relative positions and velocities of the initial clump setup.
The second and third panels show features resulting directly from the complex interplay of bow shocks surrounding the clumps. 
Note that all 10 clumps cross the same plane ($y = 0$) at roughly the same time creating a ``curtain'' or ``sheet'' of shocked material at the head of the combined bow shocks.
Note that the wings of the outer bow shocks are still visible

Throughout the later stages of its evolution run M shows a combination of bow shocks that appear as a single, global shock encompassing all of the clumps.
The formation of a single merged bow shock structure is known to occur in clumpy flows whose average separation $<d>$ is less than the lateral width of the stronger regions of the bow shock \citep{Poludnenko02}.

In the synthetic emission map of Figure~\ref{fig:Memiss}, the shock curtain (or sheet) can be seen most directly in H$\alpha$.
This occurs given the the strengthening of the shock due to the interactions among clumps.
A similar multi-clump, sheet feature is observed in HH 2 \citep{Hartigan11}.
Note also that the region of froth in run M is quite large compared to the previous simulations.
It is far more complex because some of the bright emission occurs from slower clumps that have lagged behind the fastest moving clumps but are being affected by the bow shocks from those objects.

\begin{figure}
\includegraphics[width=0.5\textwidth]{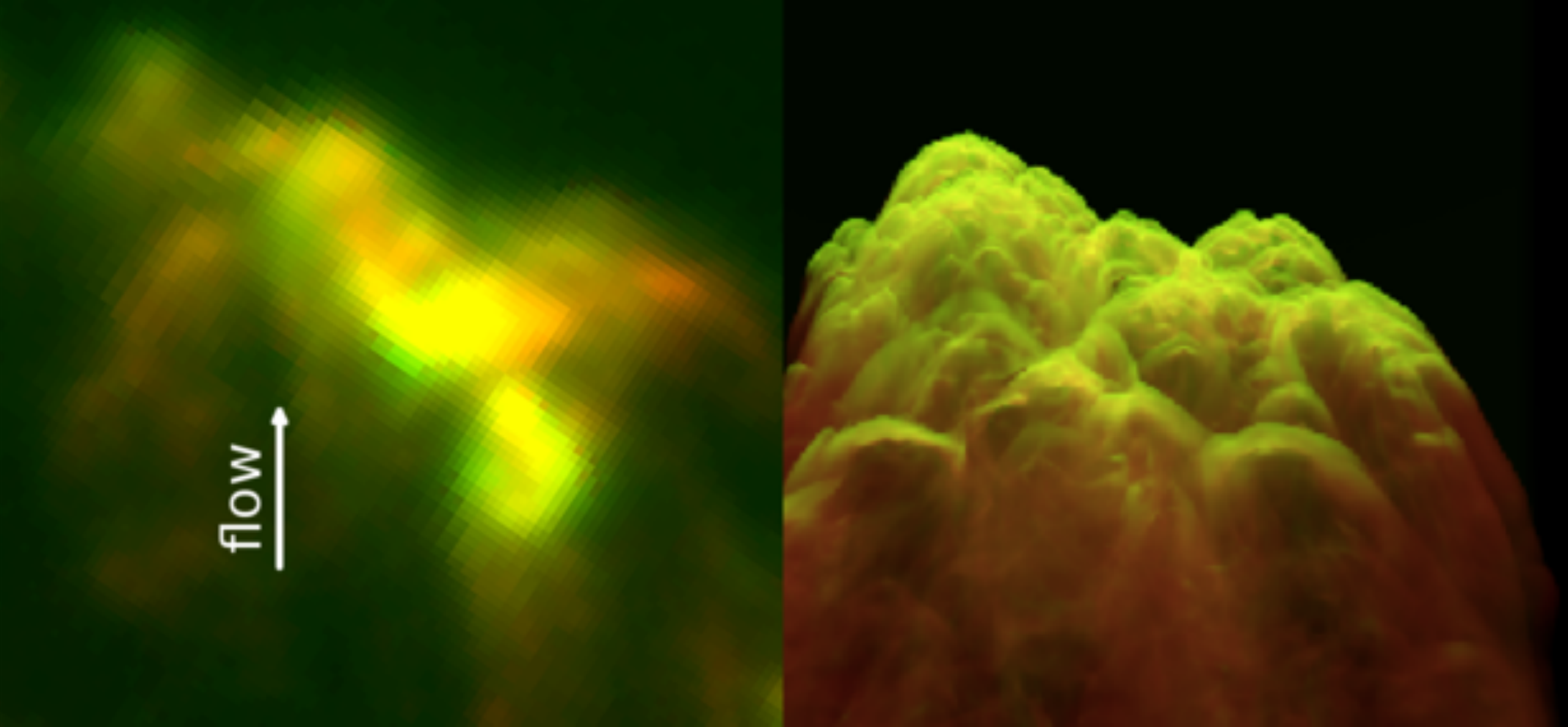}
\caption{Comparsion of HH 2 and run M.
HH 2 (\emph{left}) exhibits multiple clumps forming a shock ``sheet'' just as in simulation M (\emph{right}).
The image of HH 2 is from the Hubble Space Telescope from 1997, and this frame from simulation M was taken at a simulation time of 60 years.
Both images show the flow directed upwards.
}
\label{fig:HH2}
\end{figure}

In Figure~\ref{fig:HH2} we provide a direct comparison of the apparent shock curtain in HH 2 and synthetic emission map from run M.
The multi-epoch observations of HH 2 show the curtain evolves such that some features (clumps) are seen to pull ahead of others altering the global morphology of the structure.
This same behavior is seen in the synthetic observations as faster moving clumps push ahead within the global shock complex.
These results provide direct evidence for the inherent clumpiness of the shock curtain structures observed in HH 2, other protostellar jets and other stellar outflows such as planetary nebula and supernova remnants.

\section{Conclusions}
\label{sec:conc}
We have conducted fully 3-D radiative hydrodynamic simulations designed to study clump and bow shock interactions relevant to protostellar outflows (HH objects/jets).
Our work was specifically structured to be relevant to multi-epoch HST studies of HH jets \citep{Hartigan11}.
We used three different forms of initial conditions involving 2, 3 and 10 clumps.
Each set of initial conditions was designed to focus on a different question set associated with the dynamics or observational appearance of clumpy jets.
Within each initial condition configuration we also varied position and velocity parameters.
Synthetic emission maps of H$\alpha$ and [S II] were produced based on micro-physics calculations carried out during the simulation.  

The evolution of the clumps were consistent with previous work conducted by \citet{Poludnenko02}, \citet{Yirak10}, \citet{Yirak12}, and \citet{Klein94}.  
We find that faster moving clumps are destroyed on shorter time scales due to the faster growth times for instabilities.
In particular faster clumps have smaller cloud-crushing times $t_{cc}$, meaning transmitted shocks will compress the clump faster allowing RT and KH instabilities to begin disrupting the body of the clump on shorter timescales.
The clumps fragment due to the RT instability while material is stripped off via the KH instability.
In both cases shocked clump material becomes responsible for some of the observed emission in the post-shock regions.
The strongly cooling bow shock however dominates the emission.
NTSI modes produce strong corrugations which appear as small scale density inhomogeneities. 

It is the interaction of clump driven bow shocks with adjacent clumps (and their own bow shocks) which is the main focus of the paper. 
The interactions strongly modify the global flow producing characteristic morphological and kinematic patterns in our synthetic H$\alpha$ and [S II] emission maps.
As is well known \citep{Heathcote96}, the bow shocks are readily apparent in  H$\alpha$ while the post-shock regions are dominated by [S II] where material is rapidly cooling.
By examining maps of runs with different initial conditions (i.e. clump separation, velocity ranges), and observed at different orientations and inclination angles, we find a number of generic features in the models:

\begin{enumerate}

\item \textbf{Lateral motion of intersection points.} 
Bow shock intersection points move laterally compared to the overall direction of the flow.
These lateral motions occur as the bow shock of the faster clump sweeps over the body (and bow shock) of a slower moving clump.
Such motions have been observed in HH objects such as HH 34 \citep{Hartigan11} (see also Figure~\ref{fig:HH34}).
Thus our results indicate that such a bright H$\alpha$ spot is a shock feature rather than the result of a separate clump moving through the background jet flow.

\item \textbf{Enhanced emission at intersection points.}
Emission is enhanced at bow shock intersection points regardless of whether a Mach stem has formed or not.
The gas will either travel through a planar shock in the case of Mach reflection, or two oblique shocks in the case of regular reflection.
In both cases the material is taken to high compression ratios and produces more emission than in the case of a single bow shock.

\item \textbf{Regions of froth behind head of bow shocks.}
The term froth was first introduced by \citet{Hartigan11} when describing a filamentary emission region in HH 1 (see also Figure~\ref{fig:HH1}).
In our simulations the NTSI leads to corrugation of fast-moving bow shocks which appears in the emission as a region of strongly heterogeneous emission (froth).
Faster moving bow shocks are more susceptible to this instability and will thus show stronger density perturbations.
This translates into larger and brighter regions of H$\alpha$ froth.

\item \textbf{H$\alpha$ sheets when several bow shocks intersect.}
The combination of several clumps and bow shocks can form a continuous curtain or sheet of H$\alpha$ emission.
These sheets occur when when an array of clumps have an average spacing that is smaller than the average bow shock widths.
An example of such an H$\alpha$ curtain is observed in HH 2 \citep{Hartigan11} (see also Figure~\ref{fig:HH2}).
The presence of this shock morphology implies that ``tightly packed'' distributions of knots do occur in protostellar outflows.  
\end{enumerate}

Taken together our work supports a model of protostellar outflows where the jet beam is primarily composed of small clumps ($R_{clump} < R_{jet}$).
Thus jets are better envisioned as a spray of buckshot than a smooth beam ejected from a firehose.
As has been shown in other works, if clumps are the dominant jet component then considerations such as pulse periodicities implied by apparent knots in jet beams may be in error as random knot mergers can mimic the effect of pulsing at the jet source \citep{Raga13}.

Our work leaves open the source of the jet clumpiness.
It may be that the heterogeneity is imposed at the source or it may occur further downstream.
Laboratory astrophysics experiments which create MHD jets have shown the beams can be destabilized by m=1 kink modes \citep{Lebedev05}.
The non-linear evolution of these modes leads to well collimated clumps with a spread of velocities $\Delta V$ around the original beam velocity $V_j$.
The clumps may also form further downstream via internal working surfaces \citep{Raga09} as cooling in these shocks will tend to produce strong fragmentation.
Under the right conditions, the flow may be susceptible to cooling instabilities which is another possible mechanism for clump formation \citep{Suzuki-Vidal15}.

Finally, our work underscores the importance of multi-epoch observations.
The HST observations discussed in \citet{Hartigan11} provided a wealth of information about the behavior in astrophysical plasma systems that simply would not be available from single epoch images.
Given the emphasis placed on direct simulations in modern astrophysical studies, having time-dependent observations to inform and validate these simulations is a crucial new development.
Looking to future work, the studies in this paper were limited to purely hydrodynamic simulations.
Thus, it will be useful to introduce magnetic fields and study their effect on the observed emission.
Despite this limitation, our simulations were still able to recover many features seen in observational emission maps.

\vspace{7 mm}
\noindent\emph{Acknowledgements.} This work used the Extreme Science and Engineering Discovery Environment (XSEDE), which is supported by National Science Foundation grant number OCI-1053575.
The CIRC at the University of Rochester also provided some computational resources.
Financial support for this project was provided by the LLE at the University of Rochester, Space Telescope Science Institute grants HST-AR-11251.01-A and HST-AR-12128.01-A; by the National Science Foundation under award AST-0807363; by the Department of Energy under award de-sc0001063.

\bibliographystyle{apj}
\bibliography{mylibrary}
\end{document}